\documentclass[vecphys]{svmult}

\usepackage{makeidx}         % allows index generation
\usepackage{graphicx}        % standard LaTeX graphics tool
  \usepackage{epsfig}        % when including figure files
\usepackage{multicol}        % used for the two-column index
\usepackage{cite}            % adjusts the "syntax" of the refs in the
\usepackage[bottom]{footmisc}% places footnotes at page bottom
\makeindex             % used for the subject index
\begin{document}

\title{ARPES: A probe of electronic correlations}
\author{Riccardo Comin\inst{1} \and Andrea Damascelli\inst{1,2}}
%\institute{Department of Physics {\rm {\&}} Astronomy, University of British Columbia, Vancouver, BC, Canada
%\texttt{rcomin@physics.ubc.ca}
%\and Department of Physics {\rm {\&}} Astronomy and Quantum Matter Institute, University of British Columbia, Vancouver, BC, Canada \texttt{damascelli@physics.ubc.ca}}
\institute{Department of Physics {\rm {\&}} Astronomy, University of British Columbia, \\Vancouver, British Columbia V6T 1Z1, Canada
\and Quantum Matter Institute, University of British Columbia, \\ Vancouver, British Columbia V6T 1Z4, Canada \\
\vspace{0.25cm} Email: \texttt{rcomin@physics.ubc.ca; damascelli@physics.ubc.ca}}
% Use the package "url.sty" to avoid problems with special characters
% used in your e-mail or web address.
% Addresses should be removed from contribution and entered into
% blist.tex" (by the compiler).

\maketitle

\begin{abstract}
Angle-resolved photoemission spectroscopy (ARPES) is one of the
most direct methods of studying the electronic structure of
solids. By measuring the kinetic energy and angular distribution
of the electrons photoemitted from a sample illuminated with
sufficiently high-energy radiation, one can gain information on
both the energy and momentum of the electrons propagating inside a
material. This is of vital importance in elucidating the
connection between electronic, magnetic, and chemical structure of
solids, in particular for those complex systems which cannot be
appropriately described within the independent-particle picture.
Among the various classes of complex systems, of great interest are the transition metal oxides, which have been at the center stage in condensed matter physics for the last four decades. Following a general introduction to the topic, we will lay the theoretical basis needed to understand the pivotal role of ARPES in the study of such systems. After a brief overview on the state-of-the-art capabilities of the technique, we will review some of the most interesting and relevant case studies of the novel physics revealed by ARPES in 3\textit{d}-, 4\textit{d}- and 5\textit{d}-based oxides.
\end{abstract}

\noindent

\section{Introduction}
\label{Comin:sec:intro_correlations}

%Intro to why ARPES is useful tool to evaluate presence/role of correlations

Since their original discovery, correlated oxides\index{Transition Metal Oxides} have been extensively studied using a variety of experimental techniques and theoretical methods, thereby attracting an ever-growing interest by the community. It was soon realized that the low-energy electronic degrees of freedom were playing a key role in determining many of their unconventional properties, with the concepts of ``correlations" and ``many-body physics" gradually becoming part of the everyday dictionary of many condensed-matter physicists. It was only around the mid 90's that a series of considerable technological advancements, allowing for unprecedentedly high momentum- and energy-resolutions, made ARPES one of the prime techniques for the study of correlated materials. We will show, by discussing the required theoretical basis in conjunction with a few selected case studies, how the experimental information directly accessible using ARPES provides a unique and rich perspective towards the understanding of the electronic properties of these materials at the microscopic level.\\
%Now start talking about classes of correlated TMOs with reference to figure with periodic table
\begin{figure}[t!]
\centerline{\epsfig{file=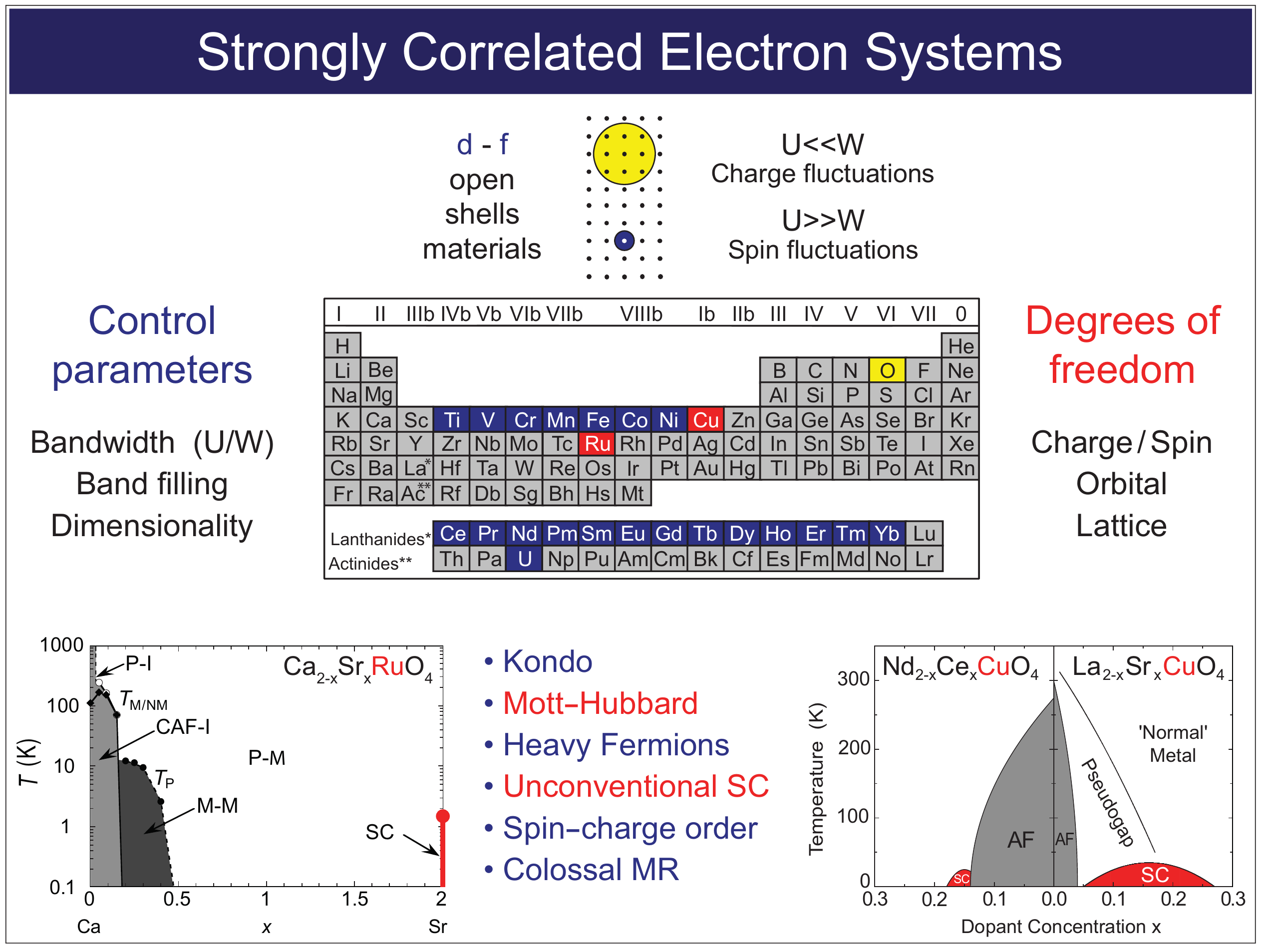,clip=,angle=0,width=1\linewidth}}
\caption{(Color). Building blocks of correlated materials and related control parameters.  Blue boxes indicate those elements exhibiting strongly localized physics ($ U \gg W$), due to the presence of localized 3\textit{d} (transition metals) and 4\textit{f} (rare earths) orbitals. Correlated physics emerges when these species form oxide compounds, and the localized \textit{d} orbitals mix with the delocalized O 2\textit{p} states. In this chapter we will review: 3\textit{d}-based materials, such as manganites, cobaltates, and cuprates, the 4\textit{d}-based ruthenates and rhodates, and the 5\textit{d}-based iridates. The phase diagrams for the special cases of Cu-based (from Ref.\,\citen{Comin:Damascelli:RMP}) and Ru-based (from Ref.\,\citen{Comin:Maeno:Ca2RuO4}) oxides, which exhibit unconventional superconductivity, are expanded in the bottom right and left panels, respectively.}
\label{Comin:Periodic_table}
\end{figure}
The correlated materials treated here belong to the ample (and growing) class of transition metals oxides (TMOs)\index{Transition Metal Oxides}. Despite giving rise to a rich variety of distinctive unconventional phenomena, the systems we will discuss all share the same basic structural elements: TM-O${}_{6}$ octahedral units, where a central transition metal (TM) cation TM${}^{n+}$ is coordinated to 6 neighbouring O${}^{2-}$ anions, sitting at positions $ {\vec{\delta}}_{i}\!=\!( \pm a, \pm a, \pm a ) $, \textit{a} being the TM-O nearest neighbour bond length. The remaining elements in the structure primarily serve the purpose of completing the stoichiometry and also, in most cases, to help controlling certain material parameters (e.g. doping, bandwidth, structure, magnetism). The delicate interplay between the \textit{localized} physics taking place \textit{within} these building blocks, and the \textit{delocalized} behaviour emerging when such local units are embedded in a crystalline matrix, is what makes these systems so complex and fascinating.

We will discuss the emerging physics in this class of materials, as one goes from the row of 3\textit{d} to that of 5\textit{d} transition metals. This is schematically illustrated in Fig.\,\ref{Comin:Periodic_table}, where the relevant elements are highlighted (see caption). The phenomenology of correlated oxides can be understood in terms of the competition between charge fluctuation (favored by the O-2\textit{p} electrons) and charge localization (driven by the TM-\textit{d} electrons). The peculiarity of 3\textit{d} and 4\textit{f} shells is that the radial part of the wave functions has an extension which is small compared to typical interatomic distances, as opposed to the oxygen 2\textit{p} orbitals which extend over many lattice sites. As a results, the localized 3\textit{d} and 4\textit{f} electrons are not well described within the independent particle picture, where electrons are assumed to interact with the average (electronic) charge density, which is hardly affected by the motion of a single electron. In reality, for the tightly confined 3\textit{d} and 4\textit{f} electrons, the addition of an extra electron in the same shell entails a large energy cost given by the strong increase in Coulomb repulsion \textit{U}. This is at the heart of what is referred to as \textit{strongly-correlated electron behavior}, and it underlies most of the spectacular phenomena observed in these materials. As a result, all the relevant degrees of freedom -- charge, spin, orbital and lattice -- are deeply entangled, and their mutual interplay is what governs the low-energy physics. 

Over time experimentalists have learned how to tune this delicate interplay by means of selected control parameters -- bandwidth, band filling, and dimensionality. All of these parameters are primarily tuned chemically (e.g. via the choice of the specific TM ion, or by carrier doping) but they can also be controlled experimentally (e.g. by means of pressure, EM fields, or in-situ doping). The resulting novel phenomena and materials include Kondo physics and heavy fermion systems (found especially in -- but not limited to -- 4\textit{f}-based materials), Mott-Hubbard/charge-transfer insulators (e.g. CuO, NiO, CoO, MnO), unconventional superconductivity (cuprates, such as e.g. La${}_{2}$CuO${}_{4}$, and also Sr${}_{2}$RuO${}_{4}$), spin-charge ordering phenomena (e.g. La${}_{2}$CoO${}_{4}$, La${}_{2}$NiO${}_{4}$, La${}_{1-x}$Ca${}_{x}$MnO${}_{3}$), and colossal magnetoresistance (La${}_{2-x}$Sr${}_{x}$Mn${}_{2}$O${}_{7}$). Remarkably, some of these phenomena can be found in the very same phase diagram, as is the case of Cu- and Ru-oxides (see bottom-right and bottom-left phase diagrams in Fig.\,\ref{Comin:Periodic_table}, respectively).

\begin{figure}[t!]
\centerline{\epsfig{file=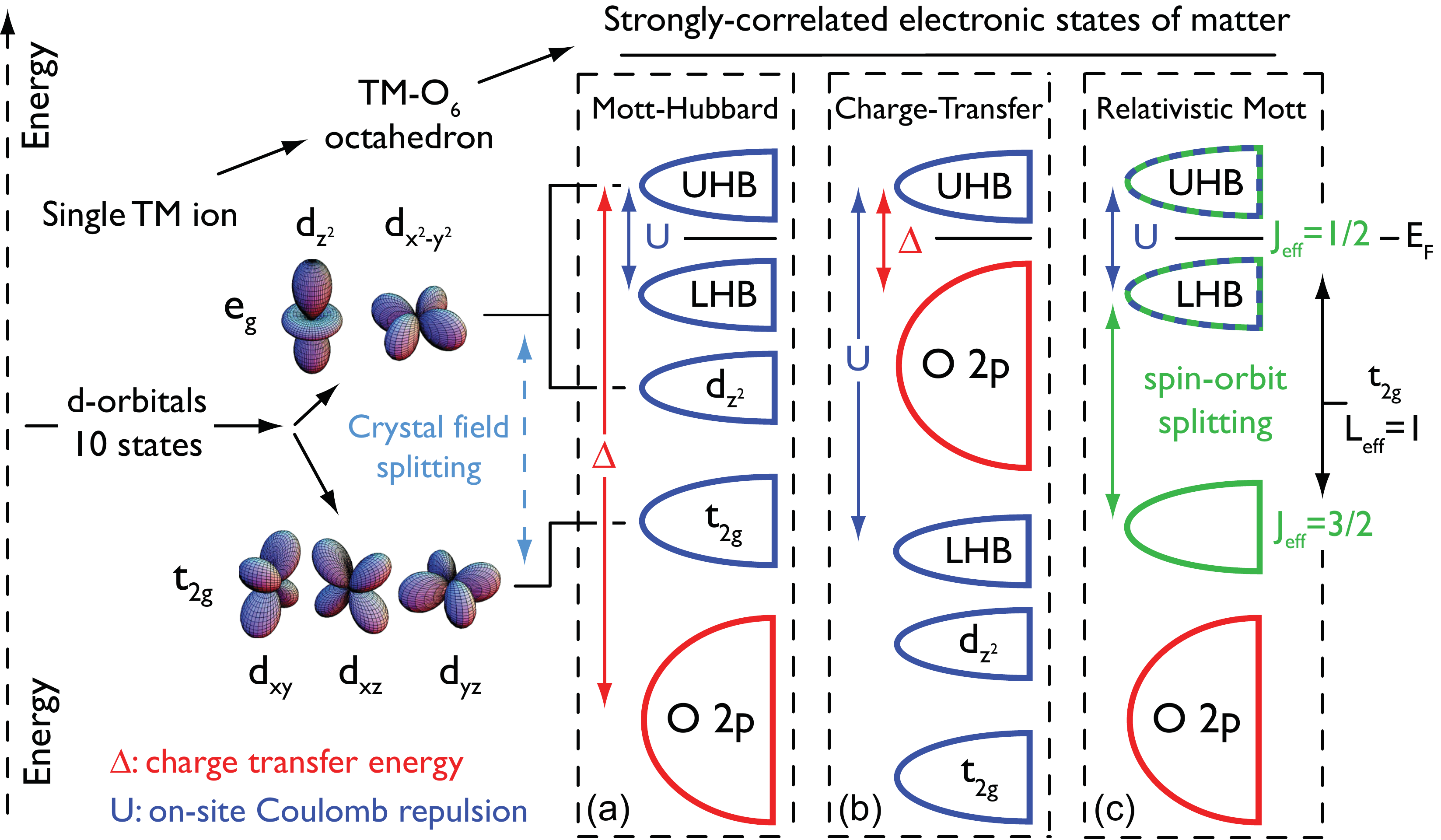,clip=,angle=0,width=1\linewidth}}
\caption{(Color). Different types of correlated quantum states of matter discovered in transition metal oxides. (a) when \textit{d}-states are close to the Fermi Energy ${E}_{F}$ (i.e., they are the \textit{lowest-ionization states}), the last half-filled \textit{d}-band is split into a lower and an upper Hubbard band by the action of \textit{U}. Whenever $ U\!>\!W $ a gap opens up at ${E}_{F}$. (b) same as (a), but now the last occupied band has O-2\textit{p} character, since \textit{U} is larger than the Cu-O charge-transfer energy $\Delta$; the corresponding gap is a \textit{charge-transfer gap}. (c) the action of \textit{spin-orbit} (SO) \textit{interaction} splits the ${t}_{2g}$ manifold into $ {J}_{eff}\!=\!3/2 $ and 1/2 submanifolds. The latter is higher in energy and lies close to ${E}_{F}$. Again, the action of \textit{U} can open a Mott-like gap, but this hinges on the previous SOI-induced splitting of the ${t}_{2g}$ band. Note: in panels (a,b) the ${d}_{{x}^{2}-{y}^{2}}$ orbital, forming the LHB and UHB, lies higher in energy than ${d}_{{z}^{2}}$ as a consequence of the tetragonal crystal field (octahedra elongated along c-axis) -- The energy levels would be inverted for compressed octahedra.}
\label{Comin:DOS_cartoon}
\end{figure}

To better elucidate the origin and nature of correlated behaviour, we will first discuss one of the most fascinating manifestations of the novel, correlated physics arising from the spatial extent of \textit{p}- and \textit{d}-orbitals: the \textit{Mott insulator}\index{Mott!insulator}\index{Insulator!Mott}. In order to understand the origin of this concept, it is useful to start from classical band theory. One of the fundamental paradigms of band theory affirms that the nature of the electronic ground state in a single-band material is entirely determined by the band filling, which is directly related to the number of electrons ${N}_{UC}$ in the unit cell: if ${N}_{UC}$ is odd (even), the system \textit{must be} metallic (insulating). For this reason, the discovery of insulating TMOs having odd ${N}_{UC}$ came as a surprise. The breakdown of single-particle physics, and consequently of band theory (where electrons are assumed to interact only with the lattice ionic potential and the average electronic density), was originally suggested by Sir. Neville Mott \cite{Comin:mott:1}. This new category of correlated insulators is the manifestation of the dominant role of on-site interactions in 3\textit{d} oxides: at half-filling (1 electron per site), the large on-site Coulomb repulsion (parametrized by the Hubbard parameter \textit{U}) between the strongly localized 3\textit{d} electrons makes hopping processes unfavourable, thus leading to charge localization and subsequent insulating behaviour, with a gap in the electronic spectrum opening up at the chemical potential. When the lowest occupied and the first unoccupied bands both have mainly \mbox{TM-\textit{d}} orbital character, as in Fig.\,\ref{Comin:DOS_cartoon}(a), we use the term \textit{Mott insulator}. The presence of a \textit{Mott gap}\index{Mott!gap}, with its characteristic scale of the order of \textit{U}, is therefore a hallmark of correlated behaviour in these systems. The lowest electron removal and addition states (bands) are respectively termed \textit{lower Hubbard band} (\textit{LHB})\index{Hubbard Band!} and \textit{upper Hubbard band} (\textit{UHB}). This is sketched in Fig.\,\ref{Comin:DOS_cartoon}(a), where a gap at the chemical potential is separating the LHB and UHB (both having mainly 3\textit{d}-character).\\
The stability of a Mott-Hubbard insulating ground state against a delocalized metallic behavior lies in the fulfilment of the \textit{Mott criterion}\index{Mott!criterion}, i.e. $ U\!>\!W $, which establishes the condition for the localizing energy scale (\textit{U}) to overcome the delocalizing ones (the bandwidth \textit{W}, proportional to intersite hopping). This criterion is based on the prerequisite that the correlated \textit{d}-states are the lowest-lying ones, i.e. those closest to ${E}_{F}$. While this is in most cases true, it fails to hold for the late 3\textit{d} transition metals \cite{Comin:ZaanenJ:Banges,Comin:Sawatzky_NiO}, where ${\epsilon}_{3d}\!<\!{\epsilon}_{2p}$ instead, ${\epsilon}$ being the orbital on-site energy (or the band center-of-mass in a delocalized picture). In such cases we talk of \textit{charge-transfer insulators} \cite{Comin:ZaanenJ:Banges}\index{Insulator!Charge-Transfer}, the denomination following from the fact that the lowest-energy excitation involves the transfer of one electron from the last occupied band, of O-2\textit{p} character, onto the first unoccupied band, of TM-3\textit{d} character. This is depicted in Fig.\,\ref{Comin:DOS_cartoon}(b), where now the \textit{charge-transfer gap} separates the 3\textit{d}-derived UHB and the O 2\textit{p}-derived valence band. A comprehensive classification can be found in Ref.\,\citen{Comin:ZaanenJ:Banges}.

The bandwidth \textit{W} and Coulomb repulsion \textit{U} are not the only relevant energy scales in the field of correlated materials. More recently, a new class of materials has appeared on the stage, that are based on the late 5\textit{d} transition metals (osmium, iridium), and whose electronic states have to be treated within a relativistic framework, due to the heavier nuclear mass. This results in a new energy scale making its way into the problem: \textit{spin-orbit} (SO) \textit{interaction}, whose strength will be indicated by ${\zeta}_{SO}$. This new element in the Hamiltonian, despite being a single-particle term (coming from the expansion of the single-fermion Dirac Hamiltonian), strongly affects the balance governing the interplay between \textit{W} and \textit{U}, making the previously introduced Mott criterion not sufficient. This results in the emergence of a new class of correlated quantum states of matter, the \textit{relativistic Mott insulator}\index{Insulator!Mott relativistic}, in which on-site Coulomb repulsion and spin-orbit interaction have to be treated on equal footing. The idea behind the existence of such a state is sketched in Fig.\,\ref{Comin:DOS_cartoon}(c). To summarize, we have introduced three different classes of correlated TMOs:
\begin{enumerate}
\item Mott-Hubbard insulators - Fig.\,\ref{Comin:DOS_cartoon}(a).
\item Charge-transfer insulators - Fig.\,\ref{Comin:DOS_cartoon}(b).
\item Relativistic Mott insulators - Fig.\,\ref{Comin:DOS_cartoon}(c).
\end{enumerate}
Please note that the 4\textit{d}-based oxides have been left out of this overview, as the presence of Mott physics in such systems is still debated, although they host a variety of different many-body phenomena, including relativistic correlated metallic behavior and unconventional superconductivity.\\
It is then clear how, as one goes down from 3\textit{d} to 5\textit{d} materials, \textit{U} progressively decreases whereas a new energy scale, spin-orbit coupling, gains importance and thus has to be accounted on equal footing. The evolution of $ U / W $ and the interplay with SO from 3\textit{d} to 5\textit{d} will be the focus of this review. In the last section of this chapter we will present a few selected examples which illustrate the different flavors of correlated electrodynamics in various TMOs (cuprates, manganites, cobaltates, ruthenates, rhodates, and iridates), and discuss the role of ARPES for quantitative and qualitative estimates of correlation effects in these systems.

\section[The ARPES technique]{The ARPES technique\footnote{Parts of the following sections have been readapted from our previous publications, Ref.\,\citen{Comin:Damascelli:RMP} and \citen{Comin:Damascelli:PS}.}}
\label{Comin:sec:introduc}

\begin{figure}[t!]
\centerline{\epsfig{figure=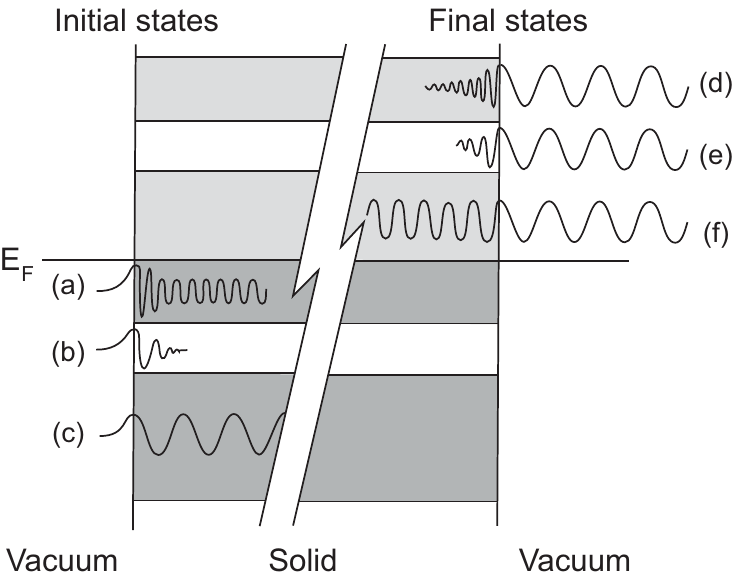,width=7cm,clip=}}
\vspace{0cm} \caption{Initial (left) and final (right) eigenstates
for the semi-infinite crystal. Left: (a) surface resonance; (b)
surface Shockley state situated in a gap of the bulk band
structure; (c) bulk Bloch state. Right: (d) surface resonance; (e)
in gap evanescent state; (f) bulk Bloch final state (from
Ref.\,\citen{Comin:meinders:1}).}\label{Comin:eigenfunctions}\end{figure}

Photoelectron spectroscopy is a general term that refers to all
those techniques based on the application of the photoelectric
effect originally observed by Hertz \cite{Comin:Hertz:1} and later
explained as a manifestation of the quantum nature of light by
Einstein \cite{Comin:EinsteinA:1}, who recognized that when light is
incident on a sample an electron can absorb a photon and escape
from the material with a maximum kinetic energy
$E_{kin}\!=\!h\nu\!-\!\phi$ (where $\nu$ is the photon frequency
and $\phi$, the material work function, is a measure of the
potential barrier at the surface that prevents the valence
electrons from escaping, and is typically 4-5 eV in metals). In
the following, we will show how the photoelectric effect also
provides us with deep insights into the quantum description of the
solid state. In particular, we will give a general overview of
angle-resolved photoemission spectroscopy (ARPES)\index{ARPES!the technique}, a highly advanced spectroscopic method that allows the direct
experimental study of the momentum-dependent electronic band
structure of solids. For a further discussion of ARPES and other
spectroscopic techniques based on the detection of photoemitted
electrons, we refer the reader to the extensive literature
available on the subject
\cite{Comin:bachrach:1,Comin:BraunJ:thea-r,Comin:brundle:1,Comin:brundle:2,Comin:cardona:1,Comin:carlsonta:1,Comin:CourthsR:Phoec.,Comin:DamascelliA:Motios,Comin:Damascelli:RMP,Comin:EastmanDE:Phosm.,Comin:FeuerbacherB:Phoesc,Comin:feuerbacher:1,Comin:Grioni:1,Comin:HimpselFJ:Ang-re,Comin:hufner:1,Comin:InglesfieldJE:Eles.,Comin:kevan:1,Comin:Leckey:1,Comin:ley:1,Comin:lindau:1,Comin:lynch:1,Comin:mahan:1,Comin:margaritondo:1,Comin:nemoshkalenko,Comin:plummer:1,Comin:ShenZX:Elesps,Comin:SmithNV:Phopm.,Comin:SmithNV:Phos.,Comin:SmithKE:elesss,Comin:wendin:1,Comin:wertheim:1,Comin:WilliamsRH:Phosst}.
\begin{figure}[t!]
\vspace{0cm} 
\begin{center}
\includegraphics*[trim=30 30 0 0,clip,width=0.85\textwidth]{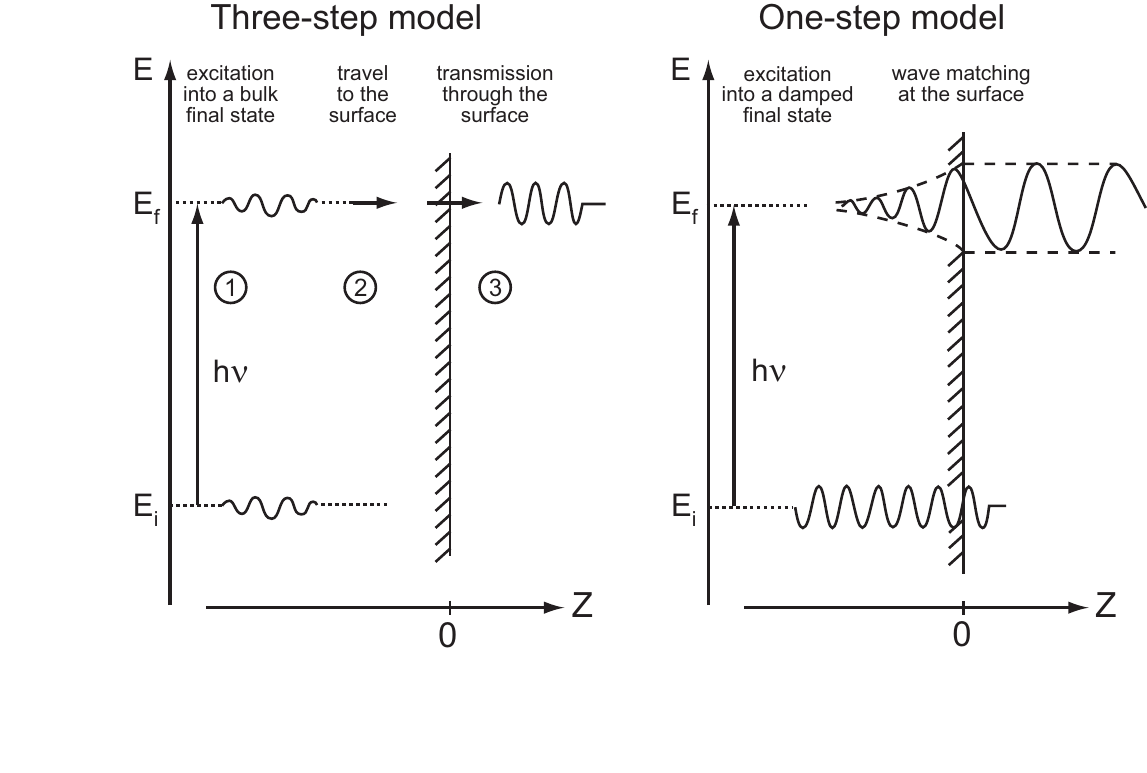}
\caption{Pictorial representation of three-step and
one-step model description of the photoemission process (from
Ref.\,\citen{Comin:hufner:1}).}
\label{Comin:1vs3step}
\end{center}
\end{figure}
As we will see in detail throughout the paper and in particular in
Sec.\,\ref{Comin:sec:sudden}, due to the complexity of the photoemission
process in solids the quantitative analysis of the experimental
data is often performed under the assumption of the {\it
independent-particle picture} and of the {\it sudden
approximation} (i.e., disregarding the many-body interactions as
well as the relaxation of the system during the photoemission
itself)\index{ARPES!Sudden approximation}. The problem is further simplified within the so-called {\it three-step model} [Fig.\,\ref{Comin:1vs3step}(a)]\index{ARPES!3-step model}, in which the photoemission event is decomposed in three independent steps:
optical excitation between the initial and final {\it bulk} Bloch
eigenstates, {\it travel} of the excited electron to the surface,
and escape of the photoelectron into vacuum after transmission
through the {\it surface} potential barrier. This is the most
common approach, in particular when photoemission spectroscopy is
used as a tool to map the electronic band structure of solids.
However, from the quantum-mechanical point of view photoemission
should not be described in terms of several independent events but
rather as a {\it one-step} process [Fig.\,\ref{Comin:1vs3step}(b)]\index{ARPES!1-step model}\index{ARPES!3-step model}: in terms of an optical transition (with probability given by Eq.\,\ref{Comin:eq:wfi}) between initial and final states consisting of
many-body wave functions that obey appropriate boundary conditions
at the surface of the solid. In particular (see Fig.\,\ref{Comin:eigenfunctions}), the initial state should be one of
the possible $N$-electron eigenstates of the semi-infinite
crystal, and the final state must be one of the eigenstates of the
ionized $(N\!-\!1)$-electron semi-infinite crystal; the latter has
also to include a component consisting of a propagating plane-wave
in vacuum (to account for the escaping photoelectron) with a
finite amplitude inside the crystal (to provide some overlap with
the initial state). 
Furthermore, as expressed by Eq.\,\ref{Comin:eq:wfi},
which represents a complete one-step description of the
problem, in order for an electron to be photoemitted in vacuum not
only there must be a finite overlap between the amplitude of
initial and final states, but the following energy and
momentum conservation laws for the impinging photon and the
$N$-electron system as a whole must also be obeyed:
\begin{eqnarray}
E_{f}^N-E_{i}^N&=&h\nu \label{Comin:eq:totenegy}
\\
{\bf k}_{f}^N-{\bf k}_{i}^N&=&{\bf k}_{h\nu}\,.
\label{Comin:eq:totmomentum}
\end{eqnarray}
\noindent
Here the indexes $i$ and $f$ refer to initial and final state,
respectively, and ${\bf k}_{h\nu}$ is the momentum of the incoming
photon. Note that, in the following, when proceeding with the more
detailed analysis of the photoemission process as well as its
application to the study of the momentum-dependent electronic
structure of solids (in terms of both conventional band mapping as
well as many-body effects), we will mainly restrict ourselves to
the context of the three-step model and the sudden approximation.

\section{Kinematics of photoemission}
\label{Comin:sec:kine}

\begin{figure}[t!]
\centerline{\epsfig{figure=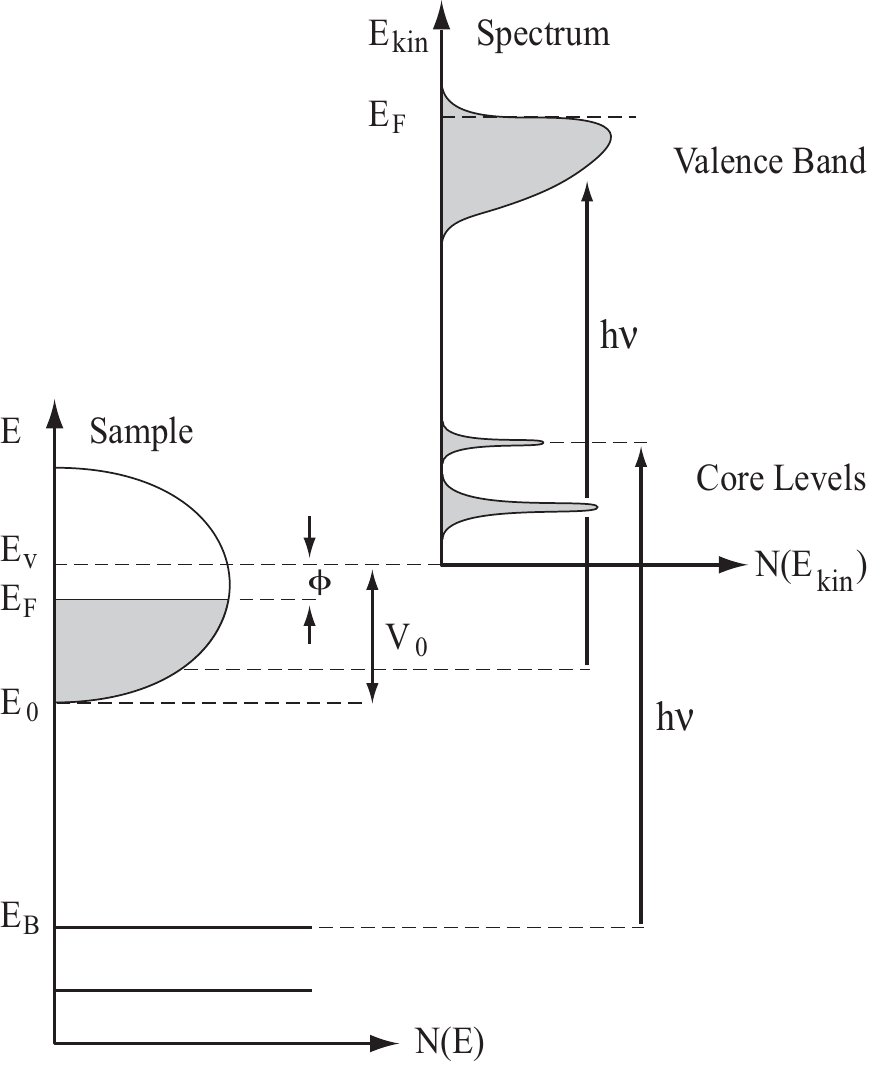,width=8cm,clip=}}
\vspace{0cm} \caption{Energetics of the photoemission process
(from Ref.\,\cite{Comin:hufner:1}). The electron energy
distribution produced by the incoming photons, and measured as a
function of the kinetic energy $E_{kin}$ of the photoelectrons
(right), is more conveniently expressed in terms of the binding
energy $E_B$ (left) when one refers to the density of states in
the solid ($E_{B}\!=\!0$ at
$E_F$).}\label{Comin:energetics}\end{figure}

\begin{figure}[t!]
\centerline{\epsfig{figure=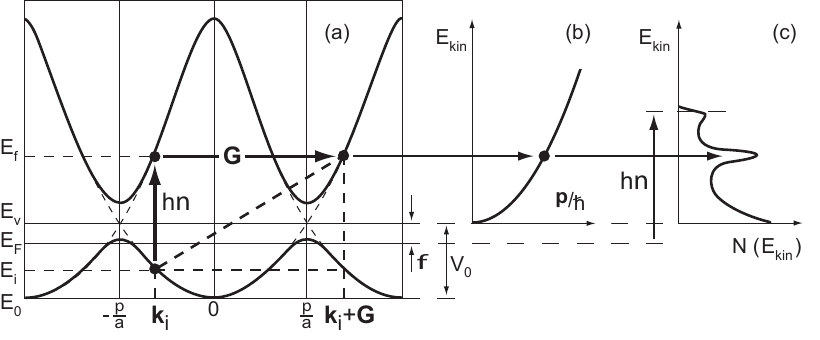,width=1\linewidth,clip=}}
\vspace{0cm} \caption{Kinematics of the photoemission\index{ARPES!Kinematics} process within the three-step nearly-free-electron final state model: (a)
direct optical transition in the solid (the lattice supplies the
required momentum, given by the \textbf{G} vector); (b) free-electron final state in vacuum; (c)
corresponding photoelectron spectrum, with a background due to the
scattered electrons ($E_{B}\!=\!0$ at $E_F$). From
Ref.\,\citen{Comin:pillo:1}.}\label{Comin:kinematics}\end{figure}

The energetics and kinematics\index{ARPES!Kinematics} of the photoemission process are shown in Fig.\ref{Comin:energetics} and\,\ref{Comin:kinematics}, while the
geometry of an ARPES experiment is sketched in
Fig.\ref{Comin:geometry}(a). A beam of monochromatized radiation supplied
either by a gas-discharge lamp, a UV laser, or a synchrotron beamline is
incident on a sample (which has to be a properly aligned single
crystal in order to perform {\it angle}- or, equivalently, {\it
momentum}-resolved measurements). As a result, electrons are
emitted by photoelectric effect and escape into the vacuum in all
directions. By collecting the photoelectrons with an electron
energy analyzer characterized by a finite acceptance angle, one
measures their kinetic energy $E_{kin}$ for a given emission
direction. This way, the wave vector or momentum ${\bf K}\!=\!{\bf
p}/\hbar$ of the photoelectrons {\it in vacuum} is also completely
determined: its modulus is given by $
K\!=\!\sqrt{2mE_{kin}}/\hbar$ and its components parallel (${\bf
K_{||}}\!=\!{K}_{x} {\hat{\bf u}}_{x} + {K}_{y} {\hat{\bf u}}_{y}$) and perpendicular (${\bf K_{\perp}}\!=\!{K}_{z} {\hat{\bf u}}_{z}$) to the sample surface are obtained in
terms of the polar ($\vartheta$) and azimuthal ($\varphi$)
emission angles defined by the experiment:
\begin{eqnarray}
K_x&=&\frac{1}{\hbar}\sqrt{2mE_{kin}}\sin\vartheta\cos\varphi
\\
K_y&=&\frac{1}{\hbar}\sqrt{2mE_{kin}}\sin\vartheta\sin\varphi
\\
K_z&=&\frac{1}{\hbar}\sqrt{2mE_{kin}}\cos\vartheta\,.
\label{Comin:eq:Kvac}
\end{eqnarray}
\noindent
The goal is then to deduce the electronic dispersion relations
$E({\bf k})$ for the solid left behind, i.e. the relation between
binding energy $E_B$ and momentum ${\bf k}$ for the electrons
propagating {\it inside} the solid, starting from $E_{kin}$ and
${\bf K}$ measured for the photoelectrons {\it in vacuum}.
In order to do that, one has to exploit the total energy and
momentum conservation laws (Eq.\,\ref{Comin:eq:totenegy}
and\,\ref{Comin:eq:totmomentum}, respectively)\index{ARPES!conservation laws}. Within the non-interacting electron picture, it is particularly straightforward to take advantage of the energy conservation law and relate, as pictorially described in Fig.\,\ref{Comin:energetics}, the kinetic energy of the photoelectron to the binding energy $E_B$ of the electronic-state inside the solid:
\begin{equation}
E_{kin}=h\nu-\phi-|E_B| \label{Comin:eq:enegy}\,.
\end{equation}
\noindent
More complex, as we will discuss below, is to gain full knowledge
of the crystal electronic momentum ${\bf k}$. Note, however, that
the photon momentum can be neglected in Eq.\,\ref{Comin:eq:totmomentum}
at the low photon energies most often used in ARPES experiments
($h\nu\!<\!100$\,eV), as it is much smaller than the typical
Brillouin-zone dimension $2\pi/a$ of a solid (see
Sec.\,\ref{Comin:sec:stateart} for more details). 
As shown in Fig.\,\ref{Comin:kinematics}, within the three-step model description
(see also Sec.\,\ref{Comin:sec:sudden}), the optical transition between
the bulk initial and final states can be described by a vertical
transition in the {\it reduced-zone scheme} (${\bf k}_{f}\!-\!{\bf
k}_{i}\!=\!0$)\index{Brillouin zone!reduced-zone scheme}, or equivalently by a transition between momentum-space points connected by a reciprocal-lattice vector
{\bf G} in the {\it extended-zone scheme} (${\bf k}_{f}\!-\!{\bf
k}_{i}\!=\!{\bf G}$)\index{Brillouin zone!extended-zone scheme}. In regard to Eq.\,\ref{Comin:eq:totenegy} and\,\ref{Comin:eq:totmomentum} and the deeper meaning of the reciprocal-lattice vector {\bf G} note that, as emphasized by
Mahan in his seminal paper on the theory of photoemission in
simple metals \cite{Comin:MahanGD:Thepsm}, ``{\it in a
nearly-free-electron gas, optical absorption may be viewed as a
two-step process. The absorption of the photon provides the
electron with the additional energy it needs to get to the excited
state. The crystal potential imparts to the electron the
additional momentum it needs to reach the excited state. This
momentum comes in multiples of the reciprocal-lattice vectors
{\boldmath$G$}. So in a reduced zone picture, the transitions are
vertical in wave-vector space. But in photoemission, it is more
useful to think in an extended-zone scheme.}'' On the contrary in
an infinite crystal with no periodic potential (i.e., a truly
free-electron gas scenario lacking of any periodic momentum
structure), no ${\bf k}$-conserving transition is possible in the
limit ${\bf k}_{h\nu}\!=\!0$, as one cannot go from an initial to
a final state along the same unperturbed free-electron parabola
without an external source of momentum. In other words, direct
transitions are prevented because of the lack of appropriate final
states (as opposed to the periodic case of
Fig.\,\ref{Comin:kinematics}).
Then again the problem would be quite different if the surface was
more realistically taken into account, as in a one-step model
description of a semi-infinite crystal. In fact, while the surface
does not perturb the translational symmetry in the $x$-$y$ plane
and ${\bf k_{\|}}$ is conserved to within a reciprocal lattice
vector ${\bf G_{\|}}$, due to the abrupt potential change along
the $z$ axis the perpendicular momentum ${\bf k}_{\bot}$ is not
conserved across the sample surface (i.e., ${\bf k_{\perp}}$ is
not a good quantum number except than deeply into the solid,
contrary to ${\bf k_{||}}$). Thus, the surface can play a direct
role in momentum conservation, delivering the necessary momentum
for indirect transitions even in absence of the crystal potential
(i.e., the so-called {\it surface photoelectric effect}\index{Surface photoelectric effect}; see also Eq.\,\ref{Comin:eq:wfi} and the related discussion).

Reverting to the three-step model {\it direct-transition}
description of Fig.\,\ref{Comin:kinematics}, the transmission through
the sample surface is obtained by matching the bulk Bloch
eigenstates inside the sample to free-electron plane waves in
vacuum. Because of the translational symmetry in the $x$-$y$ plane
across the surface, from these matching conditions it follows that
the parallel component of the electron momentum is conserved in the process:
\begin{equation}
|{\bf k}_{\|}|=|{\bf
K}_{\|}|=\frac{1}{\hbar}\sqrt{2mE_{kin}}\cdot\sin\vartheta
\label{Comin:eq:momentum}
\end{equation}
\noindent
where ${\bf k}_{\|}$ is the component parallel to the surface of
the electron crystal momentum in the {\it extended}-zone scheme (upon going to larger $\vartheta$ angles, one actually probes electrons with ${\bf k}_{\|}$ lying in higher-order Brillouin zones; by subtracting the corresponding reciprocal-lattice vector ${\bf G_{\|}}$, the {\it reduced} electron crystal momentum in the first Brillouin zone is obtained). As for the determination of ${k}_{\bot}$, which is not conserved but is also needed in order to map the electronic dispersion $E({\bf k})$ versus the total crystal wave vector ${\bf k}$, a different approach is required.
As a matter of fact, several specific experimental methods for
absolute three dimensional band mapping have been developed
\cite{Comin:hufner:1,Comin:StrocovVN:Newmab,Comin:StrocovVN:Absbmb}; however, these are rather complex and require additional and/or complementary
experimental data. Alternatively, the value of ${k}_{\bot}$
can be determined if some {\it a priori} assumption is made for
the dispersion of the electron final states involved in the
photoemission process; in particular, one can either use the
results of band structure calculations, or adopt a
nearly-free-electron description for the final bulk Bloch states:
\begin{equation}
E_f({\bf k})= \frac{\hbar^2{\bf k}^2}{2m}-|E_0|=
\frac{\hbar^2({\bf k_\|}^2+{k_\perp}^2)}{2m}-|E_0|
\label{Comin:eq:free}
\end{equation}
\noindent
where once again the electron momenta are defined in the
extended-zone scheme, and $E_0$ corresponds to the bottom of the
valence band as indicated in Fig.\,\ref{Comin:kinematics} (note that
both $E_0$ and $E_f$ are referenced to the Fermi energy $E_F$,
while $E_{kin}$ is referenced to the vacuum level $E_{v}$).
Because $E_{f}\!=\!E_{kin}\!+\!\phi$ and $\hbar^2{\bf
k}_{\|}^2/2m\!=\!E_{kin}\sin^2\vartheta$, which follow from
Fig.\,\ref{Comin:kinematics} and  Eq.\,\ref{Comin:eq:momentum}, one obtains
from Eq.\,\ref{Comin:eq:free}:
\begin{equation}
{k}_{\perp}=
\frac{1}{\hbar}\sqrt{2m(E_{kin}\cos^2\vartheta+V_0)}\,.
\label{Comin:eq:kperp}
\end{equation}
\noindent
Here $V_0\!=\!|E_0|\!+\!\phi$ is the {\it inner potential}\index{Inner potential}, which corresponds to the energy of the bottom of the valence band
referenced to vacuum level $E_{v}$. From Eq.\,\ref{Comin:eq:kperp} and
the measured values of $E_{kin}$ and $\vartheta$, if $V_0$ is also
known, one can then obtain the corresponding value of ${\bf
k}_{\perp}$. As for the determination of $V_0$, three methods are
generally used: (i) optimize the agreement between theoretical and
experimental band mapping for the occupied electronic state; (ii)
set $V_0$ equal to the theoretical zero of the muffin tin
potential used in band structure calculations; (iii) infer $V_0$
from the experimentally observed periodicity of the dispersion
$E({\bf k}_{\perp})$. The latter is actually the most convenient
method as the experiment can be realized by simply detecting the
photoelectrons emitted along the surface normal (i.e., ${\bf
K}_{\|}\!=\!0$) while varying the incident photon energy and, in
turn, the energy $E_{kin}$ of the photoelectrons and thus ${ K}_{z}$ (see Eq.\,\ref{Comin:eq:Kvac}).
%
%\begin{figure}[t!]
%\centerline{\epsfig{figure=ARPES_Intro_fig05.pdf,width=6cm,clip=}}
%\vspace{0cm} \caption{Normal and grazing emission ARPES spectra
%from Ag(100) measured with photon energies specifically chosen to
%give rise to peaks with the same binding  energy (from
%Ref.\,\cite{Comin:HansenED:Observ}).}\label{Comin:silver}\end{figure}
%
%
Note that the nearly-free electron approximation for the final
states is expected to work well for materials in which the Fermi
surface has a simple spherical (free-electron-like) topology such
as in the alkali metals, and for high-energy final states in which
case the crystal potential is a small perturbation (eventually the
final-state bands become so closely spaced in energy as to form a
continuum, and the details of the final states become
unimportant). However this approximation is also often used for
more complicated systems, even if the initial states are not free electron-like.

A particular case in which the uncertainty in ${k}_{\bot}$ is
less relevant is that of the low-dimensional systems characterized
by an anisotropic electronic structure and, in particular, a
negligible dispersion along the $z$ axis [i.e., the surface
normal, see Fig.\,\ref{Comin:geometry}(a)]. The electronic dispersion is
then almost exclusively determined by ${\bf k}_{\|}$ (as in the
case of many transition metal oxides, such as for example the
two-dimensional copper oxide superconductors
\cite{Comin:Damascelli:RMP}). As a result, one can map out in detail the
electronic dispersion relations $E({\bf k})$ simply by tracking,
as a function of ${\bf K}_{\|}$, the energy position of the peaks
detected in the ARPES spectra for different take-off angles [as in
Fig.\,\ref{Comin:geometry}(b), where both direct and inverse photoemission
spectra for a single band dispersing through the Fermi energy
$E_F$ are shown]. Furthermore, as an additional bonus associated with the lack
of a $z$ dispersion, one can directly identify the width of the
photoemission peaks as the lifetime of the photohole
\cite{Comin:SMITHNV:PHOLQL}, which contains information on the intrinsic
correlation effects of the system and is formally described by the
imaginary part of the electron self energy (see
Sec.\,\ref{Comin:sec:spectral}). On the contrary, in 3D systems the
linewidth contains contributions from both photohole and
photoelectron lifetimes, with the latter reflecting final state
scattering processes and thus the finite probing depth; as a
consequence, isolating the intrinsic many-body effects becomes a
much more complicated problem.

\section{Three-step model and sudden approximation}
\label{Comin:sec:sudden}

\index{ARPES!3-step model}\index{ARPES!Sudden approximation}
To develop a formal description of the photoemission process, one
has to calculate the transition probability $w_{fi}$ for an
optical excitation between the $N$-electron ground state
$\Psi_i^N$ and one of the possible final states $\Psi_f^N$. This
can be approximated by Fermi's golden rule:
\begin{equation}
w_{fi}=\frac{2\pi}{\hbar}|\langle\Psi_f^N|H_{int}|\Psi_i^N\rangle|^2\delta(E_f^N-E_i^N-h\nu)
\label{Comin:eq:wfi}
\end{equation}
\noindent
where $E_i^N\!=\!E_i^{N-1}\!-\!E_B^{\bf k}$ and
$E_f^N\!=\!E_f^{N-1}\!+\!E_{kin}$ are the initial and final-state
energies of the $N$-particle system ($E_B^{\bf k}$ is the binding
energy of the photoelectron with kinetic energy $E_{kin}$ and
momentum ${\bf k}$). The interaction with the photon is treated as
a perturbation given by:
\begin{equation}
H_{int}=\frac{e}{2mc}({\bf A}\!\cdot\!{\bf p}+{\bf p}\!\cdot\!{\bf
A})= \frac{e}{mc}\,{\bf A}\!\cdot\!{\bf p} \label{Comin:eq:hint}
\end{equation}
\noindent
where {\bf p} is the electronic momentum operator and {\bf A} is
the electromagnetic vector potential (note that the gauge
$\Phi\!=\!0$ was chosen for the scalar potential $\Phi$, and the
quadratic term in ${\bf A}$ was dropped because in the linear
optical regime it is typically negligible with respect to the
linear terms). In Eq.\,\ref{Comin:eq:hint} we also made use of the
commutator relation $[{\bf p},{\bf
A}]\!=\!-i\hbar\nabla\!\cdot\!{\bf A}$ and dipole approximation
[i.e., {\bf A} constant over atomic dimensions and therefore
$\nabla\!\cdot\!{\bf A}\!=\!0$, which holds in the ultraviolet].
Although this is a routinely used approximation, it should be
noted that $\nabla\!\cdot\!{\bf A}$ might become important at the
{\it surface} where the electromagnetic fields may have a strong
spatial dependence. This surface photoemission contribution, which
is proportional to ($\varepsilon-1$) where $\varepsilon$ is the
medium dielectric function, can interfere with the bulk
contribution resulting in asymmetric lineshapes for the bulk
direct-transition peaks
\cite{Comin:feuerbacher:1,Comin:MillerT:Intbbs,Comin:HansenED:SurpA,Comin:HansenED:Ovetsp}.
At this point, a more rigorous approach is to proceed with the
so-called {\it one-step model} [Fig.\,\ref{Comin:1vs3step}(b)], in which
photon absorption, electron removal, and electron detection are
treated as a single coherent process
\cite{Comin:mitchell:1,Comin:makinson:1,Comin:buckingham:1,Comin:MahanGD:Thepsm,Comin:SchaichWL:Modctt,Comin:FeibelmanPJ:Phos-C,Comin:PendryJB:Thep.,Comin:pendry:1,Comin:LiebschA:Thepla,Comin:liebsch:1,Comin:LindroosM:Sursa-,Comin:LindroosM:novdmF,Comin:BansilA:Ang-re,Comin:BansilA:Mateet,Comin:BansilA:Impmet}.
In this case bulk, surface, and vacuum have to be included in the
Hamiltonian describing the crystal, which implies that not only
bulk states have to be considered but also surface and evanescent
states, as well as surface resonances (see Fig.\,\ref{Comin:eigenfunctions}).
Note that, under the assumption $\nabla\!\cdot\!{\bf A}\!=\!0$,
from Eq.\,\ref{Comin:eq:hint} and the commutation relation $[H_0,{\bf
p}]\!=\!i\hbar \nabla V$ (where $H_0\!=\!{\bf p}^2/2m\!+\!V$ is
the unperturbed Hamiltonian of the semi-infinite crystal) it
follows that the matrix elements appearing in Eq.\,\ref{Comin:eq:wfi}
are proportional to $\langle\Psi_f^N|{\bf A}\!\cdot\!\nabla
V|\Psi_i^N\rangle$. This explicitly shows that for a true
free-electron like system it would be impossible to satisfy
simultaneously energy and momentum conservation laws inside the
material because there $\nabla V\!=\!0$. The only region where
electrons could be photoexcited is at the surface where $\partial
V/\partial z\!\neq\!0$, which gives rise to the so-called {\it
surface photoelectric effect}\index{Surface photoelectric effect}. However, due to the complexity of the one-step model, photoemission data are usually discussed
within the three-step model [Fig.\,\ref{Comin:1vs3step}(a)], which,
although purely phenomenological, has proven to be rather
successful \cite{Comin:FanHY:1,Comin:berglund:1,Comin:FeibelmanPJ:Phos-C}. Within
this approach, the photoemission process is subdivided into three
independent and sequential steps:
\begin{figure}[t!]
\centerline{\epsfig{figure=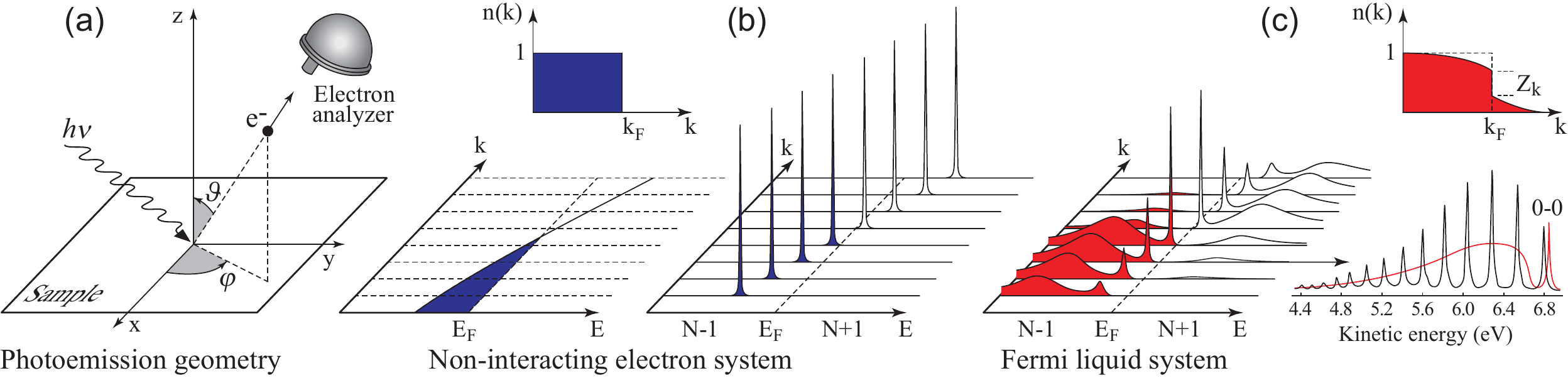,width=1\linewidth,clip=}}
\caption{(a) Geometry of an ARPES experiment; the emission
direction of the photoelectron is specified by the polar
($\vartheta$) and azimuthal ($\varphi$) angles. Momentum resolved
one-electron removal and addition spectra for: (b) a
non-interacting electron system (with a single energy band
dispersing across the Fermi level); (c) an interacting Fermi
liquid system. The corresponding ground-state ($T\!=\!0$ K)
momentum distribution function $n({\bf k})$ is also shown. (c)
Bottom right: photoelectron spectrum of gaseous hydrogen (black) and the ARPES spectrum of solid hydrogen developed from the gaseous one (red). Adapted from Ref.\,\citen{Comin:Damascelli:RMP}.
}\label{Comin:geometry}\end{figure}
\begin{itemize}
\item[(i)]{\hspace{0.2cm}Optical excitation of the electron in the {\textit{bulk}}.}
\item[(ii)]{\hspace{0.2cm}Travel of the excited electron to the surface.}
\item[(iii)]{\hspace{0.2cm}Escape of the photoelectron into vacuum.}
\end{itemize}
 The total photoemission intensity is then given by the product
of three independent terms: the total probability for the optical
transition, the scattering probability for the travelling
electrons, and the transmission probability through the surface
potential barrier. Step (i) contains all the information about the
intrinsic electronic structure of the material and will be
discussed in detail below. Step (ii) can be described in terms of
an effective mean free path, proportional to the probability that
the excited electron will reach the surface without scattering
(i.e, with no change in energy and momentum). The inelastic
scattering processes, which determine the surface sensitivity of
photoemission (see Sec.\,\ref{Comin:sec:stateart}), give rise to a
continuous background in the spectra which is usually ignored or
subtracted. Step (iii) is described by a transmission probability
through the surface, which depends on the energy of the excited
electron and the material work function $\phi$ (in order to have
any finite escape probability the condition $\hbar^2 {k}_{\perp}^2/2m\!\geq\!|E_0|\!+\!\phi$ must be satisfied).

In evaluating step (i), and therefore the photoemission intensity
in terms of the transition probability $w_{fi}$, it would be
convenient to factorize the wavefunctions in Eq.\,\ref{Comin:eq:wfi}
into photoelectron and $(N\!-\!1)$-electron terms, as we have done
for the corresponding energies. The final state $\Psi^{N}_{f}$ then becomes:
\begin{equation}
\Psi_f^N={\mathcal{A}}\, \phi_f^{\bf k}\, \Psi^{N-1}_f
\end{equation}
\noindent
where $\mathcal{A}$ is an antisymmetric operator that properly
antisymmetrizes the $N$-electron wavefunction so that the Pauli
principle is satisfied, $\phi_f^{\bf k}$ is the wavefunction of
the photoelectron with momentum ${\bf k}$, and $\Psi^{N-1}_{f}$ is
the final state wavefunction of the $(N\!-\!1)$-electron system
left behind, which can be chosen as an excited state with
eigenfunction $\Psi^{N-1}_{m}$ and energy $E^{N-1}_m$. The {\it
total} transition probability is then given by the sum over {\it
all} possible excited states $m$. This derivation, which originated from writing the transition probability using Fermi's golden rule, Eq.\,\ref{Comin:eq:wfi}, implicitly assumes the validity of the so-called \textit{sudden approximation}, which is extensively used in many-body calculations of the photoemission spectra from interacting electron systems, and is in principle applicable only to high kinetic-energy electrons. In this limit, the photoemission process is assumed to be {\it sudden}, with no post-collisional interaction between the photoelectron and the system left behind (in other words, an electron is instantaneously removed and the effective potential of the system changes discontinuously at that instant)\footnote{In particular, this implies that the wavefunction for the (N-1)-electron system at time $ {t}_{1} $ (when the interaction $ {H}_{int} $ is switched on) remains unchanged when $ {H}_{int} $ is switched off at $ {t}_{2} $, thus allowing to use Fermi's golden rule and the instantaneous transition probabilities ${w}_{fi}$. This approximation is only valid when $ {t}_{2} - {t}_{1} \ll \frac{\hbar}{\Delta E} $, $ {\Delta E} $ being the characteristic energy separation of the (N-1) system \cite{Comin:Sakurai}.}. Note, however, that the sudden approximation is inappropriate for low kinetic energy photoelectrons, which may need longer than the system response
time to escape into vacuum. In this case, the so-called {\it
adiabatic limit}\index{ARPES!Adiabatic limit}, one can no longer use the instantaneous transition amplitudes $ {w}_{fi} $ and the detailed screening of photoelectron and photohole has to be taken into account \cite{Comin:GadzukJW:Excedc}.

For the initial state, let us first assume for simplicity that
$\Psi^{N}_{i}$ is a single Slater determinant (i.e., Hartree-Fock
formalism), so that we can write it as the product of a
one-electron orbital $\phi_i^{\bf k}$ and an $(N\!-\!1)$-particle
term:
\begin{equation}
\Psi_i^N={\mathcal{A}}\, \phi_i^{\bf k}\, \Psi^{N-1}_i\,.
\end{equation}
\noindent
More generally, however, $\Psi^{N-1}_{i}$ should be expressed as
$\Psi^{N-1}_i\!=\!c_{\bf k}\Psi^{N}_i$, where $c_{\bf k}$ is the
annihilation operator for an electron with momentum $\bf k$. This
also shows that $\Psi^{N-1}_{i}$ is {\it not} an eigenstate of the
$(N\!-\!1)$ particle Hamiltonian, but is just what remains of the
$N$-particle wavefunction after having pulled out one electron. At
this point, we can write the matrix elements in Eq.\,\ref{Comin:eq:wfi}
as:
\begin{equation}
\langle\Psi_f^N|H_{int}|\Psi_i^N\rangle\!=\!\langle\phi_f^{\bf
k}|H_{int}|\phi_i^{\bf k}\rangle
\langle\Psi^{N-1}_{m}|\Psi^{N-1}_{i}\rangle
\end{equation}
\noindent
where $\langle\phi_f^{\bf k}|H_{int}|\phi_i^{\bf
k}\rangle\!\equiv\!M_{f,i}^{\bf k}$ is the one-electron dipole
matrix element, and the second term is the $(N\!-\!1)$-electron
overlap integral. Here, we replaced ${\Psi}_{f}^{N-1}$ with an
eigenstate ${\Psi}_{m}^{N-1}$, as discussed above. The total
photoemission intensity measured as a function of $E_{kin}$ at a
momentum ${\bf k}$, namely $I({\bf
k},E_{kin})\!=\!\sum_{f,i}w_{f,i}$, is then proportional to:
\begin{equation}
\sum_{f,i}|M_{f,i}^{\bf k}|^2 \sum_{m}
|c_{m,i}|^2\delta(E_{kin}\!+\!E_{m}^{N-1}\!-E_i^N\!-\!h\nu)
\label{Comin:eq:photocurrent}
\end{equation}
\noindent
where $|c_{m,i}|^2\!=\!|\langle\Psi^{N-1}_{m}|\Psi^{N-1}_{i}\rangle|^2$
is the probability that the removal of an electron from state $i$
will leave the $(N\!-\!1)$-particle system in the excited state
$m$. From here we see that, if
$\Psi^{N-1}_{i}\!=\!\Psi^{N-1}_{m_0}$ for one particular
$m\!=\!m_0$, the corresponding $|c_{m_0,i}|^2$ will be unity and
all the others $c_{m,i}$ zero; in this case, if also $M_{f,i}^{\bf
k}\!\neq\!0$, the ARPES spectra will be given by a delta function
at the Hartree-Fock orbital energy $E_B^{\bf k}\!=\!-\varepsilon_{\bf
k}^{b}$, as shown in Fig.\,\ref{Comin:geometry}(b) (i.e., non-interacting
particle picture). In the strongly correlated systems, however,
many of the $|c_{m,i}|^2$ terms will be different from zero because the
removal of the photoelectron results in a strong change of the
system effective potential and, in turn, $\Psi^{N-1}_{i}$ will
have an overlap with many of the eigenstates $\Psi^{N-1}_{m}$.
Therefore, the ARPES spectra will not consist of single delta
functions but will show a main line and several satellites
according to the number of excited states $m$ created in the
process [Fig.\,\ref{Comin:geometry}(c)].

\section{One-particle spectral function} \label{Comin:sec:spectral}
\index{Spectral function}

In the discussion of photoemission on solids, and in particular on
the correlated electron systems in which many $|c_{m,i}|^2$ in
Eq.\,\ref{Comin:eq:photocurrent} are different from zero, the most
powerful and commonly used approach is based on the Green's
function formalism
\cite{Comin:Abrikosov:1,Comin:hedin:1,Comin:FetterAL:1,Comin:mahan:2,Comin:economou:1,Comin:rickayzen:1}. In this context, the propagation of a single electron in a many-body
system is described by the {\it time-ordered} one-electron Green's
function ${\mathcal{G}}({\bf k},t-t^\prime)$\index{Green's function!time-ordered}, which can be interpreted as the probability amplitude that an electron added to the system in a Bloch state with momentum ${\bf k}$ at a time zero will still be in the same state after a time $|t\!-\!t^\prime|$. By taking the
Fourier transform, ${\mathcal{G}}({\bf k},t\!-\!t^\prime)$ can be
expressed in energy-momentum representation resulting in
${\mathcal{G}}({\bf k},\omega)\!=\!G^+({\bf
k},\omega)+G^-({\bf k},\omega)$, where $G^+({\bf k},\omega)$
and $G^-({\bf k},\omega)$ are the one-electron addition and
removal Green's function\index{Green's function!electron addition}\index{Green's function!electron removal}, respectively. At $T\!=\!0$:
\begin{equation}
G^\pm({\bf
k},\omega)=\sum_{m}\frac{|\langle\Psi^{N\pm1}_m|c^\pm_{\bf k}
|\Psi^N_i\rangle|^2}{\omega- E_m^{N\pm 1}+ E^N_i\pm i\eta}
\end{equation}
\noindent
where the operator $c_{\bf k}^+\!=\!c_{{\bf k}\sigma}^\dag$
($c_{\bf k}^-\!=\!c_{{\bf k}\sigma}^{ }$) creates (annihilates) an
electron with energy $\omega$, momentum ${\bf k}$, and spin
$\sigma$ in the $N$-particle initial state $\Psi^N_i$; the
summation runs over all possible $({N\!\pm\!1})$-particle
eigenstates $\Psi_m^{N\pm1}$ with eigenvalues $E_m^{N\pm1}$, and
$\eta$ is a positive infinitesimal (note also that from here on we
will take $\hbar\!=\!1$). In the limit $\eta\!\rightarrow\!0^+$
one can make use of the identity
\mbox{$(x\!\pm\!i\eta)^{-1}\!=\!{\mathcal{P}}(1/x)\!\mp\!i\pi\delta(x)$},
where $\mathcal{P}$ denotes the principle value, to obtain the
{\it one-particle spectral function}  $A({\bf
k},\omega)\!=\!A^+({\bf k},\omega)\!+\!A^-({\bf
k},\omega)\!=\!-(1/\pi){\rm Im}\,G({\bf k},\omega)$, with:
\begin{equation}
A^\pm({\bf
k},\omega)\!=\!\!\!\sum_{m}|\langle\Psi^{N\pm1}_m\!|c^\pm_{\bf k}
|\Psi^N_i\rangle|^2 \delta(\omega\!-\!E_m^{N\pm 1}\!\!+\!E^N_i)
\label{Comin:eq:akw0}
\end{equation}
\noindent
and $G({\bf k},\omega)\!=\!G^+({\bf k},\omega)\!+\![G^{-}({\bf
k},\omega)]^*$, which defines the {\it retarded} Green's function\index{Green's function!retarded}.
Note that $A^-({\bf k},\omega)$ and $A^+({\bf k},\omega)$ define
the one-electron removal and addition spectra which one can probe
with direct and inverse photoemission, respectively. This is
evidenced, for the direct case, by the comparison between the
expression for $A^-({\bf k},\omega)$ and
Eq.\,\ref{Comin:eq:photocurrent}\index{ARPES!Cross-section} for the photoemission intensity (note that in the latter $\Psi^{N-1}_i\!=\!c_{\bf k}\Psi^{N}_i$ and the
energetics of the photoemission process has been explicitly
accounted for). Finite temperatures effect can be taken into
account by extending the Green's function formalism just
introduced to $T\!\neq\!0$ (see, e.g.,
Ref.\,\cite{Comin:mahan:2}). In the latter case, by invoking once
again the sudden approximation, the intensity measured in an ARPES
experiment on a 2D  single-band system can be conveniently written
as:
\begin{equation}
I({\bf k},\omega)=I_0({\bf k},\nu,{\bf A})f(\omega)A({\bf
k},\omega) \label{Comin:eq:ikw} \label{Comin:eq:intone}
\end{equation}
\noindent
where ${\bf k}\!=\!{\bf k}_\parallel$ is the in-plane electron
momentum, $\omega$ is the electron energy with respect to the
Fermi level, and $I_0({\bf k},\nu,{\bf A})$ is proportional to the
squared one-electron matrix element $|M_{f,i}^{\bf k}|^2$ and
therefore depends on the electron momentum, and on the energy and
polarization of the incoming photon. We also introduced the Fermi
function\index{Fermi function} $f(\omega)\!=\!(e^{\omega/k_BT}\!+\!1)^{-1}$,  which accounts for the fact that direct photoemission probes only the
occupied electronic states. Note that in Eq.\,\ref{Comin:eq:ikw} we
neglected the presence of any extrinsic background and the
broadening due to the energy and momentum resolution, which
however have to be carefully considered when performing a
quantitative analysis of the ARPES spectra (see
Sec.\,\ref{Comin:sec:matele} and Eq.\,\ref{Comin:eq:ikwreal}).

The corrections to the Green's function originating from
electron-electron correlations can be conveniently expressed in
terms of the electron {\it proper self energy} $\Sigma({\bf
k},\omega)\!=\!\Sigma^{\prime}({\bf
k},\omega)\!+\!i\Sigma^{\prime\prime}({\bf k},\omega)$\index{Self-energy}. Its real and imaginary parts contain all the information on the energy
renormalization and lifetime, respectively, of an electron with
band energy ${\varepsilon}_{k}^{b}$ and momentum ${\bf k}$ propagating in a
many-body system. The Green's and spectral functions expressed in
terms of the self energy are then given by:
\begin{equation}
G({\bf k},\omega)=\frac{1}{\omega-\varepsilon_{\bf k}^{b}-\Sigma({\bf
k},\omega)} \label{Comin:eq:gwk}
\end{equation}
%
%\vspace{-.55cm}
\begin{equation} A({\bf
k},\omega)=-\frac{1}{\pi}\frac{\Sigma^{\prime\prime}({\bf
k},\omega)}{[\omega-\varepsilon_{\bf k}^{b}-\Sigma^{\prime}({\bf
k},\omega)]^2+[\Sigma^{\prime\prime}({\bf k},\omega)]^2}\,.
\label{Comin:eq:akw1}
\end{equation}
\noindent
Because $G(t,t^\prime)$ is a linear response function to an
external perturbation, the real and imaginary parts of its Fourier
transform $G({\bf k},\omega)$ have to satisfy causality and,
therefore, also Kramers-Kronig relations. This implies
that if the full $A({\bf k},\omega)\!=\!-(1/\pi){\rm Im}\,G({\bf
k},\omega)$ is available from photoemission and inverse
photoemission, one can calculate ${\rm Re}\,G({\bf k},\omega)$ and
then obtain both the real and imaginary parts of the self energy
directly from Eq.\,\ref{Comin:eq:gwk}. However, due to the lack of
high-quality inverse photoemission data, this analysis is usually
performed using only ARPES spectra by taking advantage of certain
approximations (such as, e.g., particle-hole symmetry near $E_F$; for a more detailed discussion, see also Ref.\,\citen{Comin:VeenstraCN:elusive_elph,Comin:VeenstraCN:SF_Holstein} and references therein).

In general, the exact calculation of $\Sigma({\bf k},\omega)$ and,
in turn, of $A({\bf k},\omega)$ is an extremely difficult task. In
the following, as an example we will briefly consider the
interacting FL case \cite{Comin:landau:1,Comin:landau:2,Comin:landau:3}\index{Self-energy!Fermi liquid}\index{Spectral function!Fermi liquid}. Let us start from the trivial $\Sigma({\bf k},\omega)\!=\!0$
non-interacting case.  The $N$-particle eigenfunction $\Psi^N$ is
a single Slater determinant and we always end up in a single
eigenstate when removing or adding an electron with momentum ${\bf
k}$. Therefore, $G({\bf
k},\omega)\!=\!1/(\omega\!-\!\varepsilon_k^b\!\pm\!i\eta)$ has only one
pole for each ${\bf k}$, and $A({\bf
k},\omega)\!=\!\delta(\omega\!-\!\varepsilon_k^b)$ consists of a single
line at the band energy $\varepsilon_k^b$ [as shown in
Fig.\,\ref{Comin:geometry}(b)]. In this case, the occupation numbers
$n_{{\bf k}\sigma}^{ }\!=\!c_{{\bf k}\sigma}^\dag c_{{\bf
k}\sigma}^{ }$ are good quantum numbers and for a metallic system
the {\it momentum distribution} [i.e., the expectation value
$n({\bf k})\!\equiv\!\langle n_{{\bf k}\sigma}\rangle$, quite
generally independent of the spin $\sigma$ for nonmagnetic
systems], is characterized by a sudden drop from 1 to 0 at ${\bf
k}\!=\!{\bf k}_F$ [Fig.\,\ref{Comin:geometry}(b), top], which defines a
sharp Fermi surface. If we now switch on the electron-electron
correlation adiabatically, (so that the system remains at
equilibrium), any particle added into a Bloch state has a certain
probability of being scattered out of it by a collision with
another electron, leaving the system in an excited state in which
additional electron-hole pairs have been created. The momentum
distribution $n({\bf k})$ will now show a discontinuity smaller
than 1 at ${\bf k}_F$ and a finite occupation probability for
${\bf k}\!>\!{\bf k}_F$ even at $T\!=\!0$ [Fig.\,\ref{Comin:geometry}(c),
top]. As long as $n({\bf k})$ shows a finite discontinuity $Z_{\bf
k}\!>\!0$ at ${\bf k}\!=\!{\bf k}_F$, we can describe the
correlated Fermi sea in terms of well defined {\it
quasiparticles}\index{Quasiparticle}, i.e. electrons {\it dressed} with a manifold of excited states, which are characterized by a pole structure
similar to the one of the non-interacting system but with
renormalized energy $\varepsilon_{\bf k}^q$, mass $m^*$, and a
finite lifetime $\tau_{\bf k}\!=\!1/\Gamma_{\bf k}$. In other
words, the properties of a FL are similar to those of a free
electron gas with damped quasiparticles. As the bare-electron
character of the quasiparticle or pole strength (also called
coherence factor) is $Z_{\bf k}\!<\!1$ and the total spectral
weight must be conserved (see Eq.\,\ref{Comin:eq:intakw}), we can
separate $G({\bf k},\omega)$ and $A({\bf k},\omega)$ into a {\it
coherent} pole part and an {\it incoherent} smooth part without
poles \cite{Comin:pines:1}:
\begin{equation}
G({\bf k},\omega)=\frac{Z_{\bf k}}{\omega-\varepsilon_{\bf
k}^{q}+i\Gamma_{\bf k}}+G_{incoh} \label{Comin:eq:gfl}
\end{equation}
%
%\vspace{-.55cm}
\begin{equation} A({\bf
k},\omega)=Z_{\bf k}\frac{\Gamma_{\bf k}/\pi}
{(\omega-\varepsilon_{\bf k}^{q})^2+\Gamma_{\bf k}^2}+A_{incoh}
\end{equation}
\noindent
where $Z_{\bf k}\!=\!(1\!-\!\frac{\partial\Sigma^{\prime}}{\partial\omega})^{-1}$,
$\varepsilon_{\bf k}^q\!=\!Z_{\bf k}(\varepsilon_{\bf
k}^b\!+\!\Sigma^{\prime})$, $\Gamma_{\bf k}\!=\!Z_{\bf
k}|\Sigma^{\prime\prime}|$, and the self energy and its
derivatives are evaluated at $\omega\!=\!\varepsilon_{\bf k}^q$. It
should be emphasized that the FL description is valid only in
proximity to the Fermi surface and rests on the condition
$\varepsilon_{\bf k}^q\!-\!\mu\!\gg\!|\Sigma^{\prime\prime}|$ for
small $(\omega\!-\!\mu)$ and $({\bf k}\!-\!{\bf k}_F)$\index{Self-energy}\index{Spectral function}.
Furthermore, $\Gamma_{\bf k}\!\propto\![(\pi
k_BT)^2\!+\!(\varepsilon_{\bf k}^q\!-\!\mu)^2]$ for a FL system in
two or more dimensions \cite{Comin:luttinger:2,Comin:pines:1}, although
additional logarithmic corrections should be included in the
two-dimensional case \cite{Comin:HodgesC:EffFsg}. By comparing the
electron removal and addition spectra for a FL of quasiparticles
with those of a non-interacting electron system (in the lattice
periodic potential), the effect of the self-energy correction
becomes evident [see Fig.\,\ref{Comin:geometry}(c) and \,\ref{Comin:geometry}(b), respectively]. The quasiparticle\index{Quasiparticle} peak has now a finite lifetime and width (due to $\Sigma^{\prime\prime}$), but sharpens rapidly as it emerges from the broad incoherent component and approaches the Fermi level, where the lifetime is infinite corresponding to a
well defined quasiparticle [note that the coherent and incoherent
part of $A({\bf k},\omega)$ represent the main line and satellite
structure discussed in the previous section and shown in
Fig.\,\ref{Comin:geometry}(c), bottom right].
Furthermore, the peak position is shifted with respect to the bare
band energy $\varepsilon_{\bf k}^b$ (due to $\Sigma^{\prime}$): as the
quasiparticle mass is larger than the band mass because of the
dressing ($m^*\!>\!m$), the total dispersion (or bandwidth) will
be smaller ($|\varepsilon_{\bf k}^q|\!<\!|\varepsilon_{\bf k}^b|$). We note here, as later discussed in more detail in relation to Fig.\,\ref{Comin:Polaron}, that the continuum of excitations described by the incoherent part of $A({\bf k},\omega)$ in general does still retain a \textbf{k} and $ \omega $-dependent structure with spectral weight distributed predominately along the non-interacting bare band; this, however, is usually characterized by remarkably broad lineshapes [see e.g. Fig.\,\ref{Comin:H2_model}(c) and \ref{Comin:Manganites}] and should not be mistaken for a quasiparticle dispersion.

Among the general properties of the spectral function there are
also several sum rules. A fundamental one, which in discussing the
FL model was implicitly used to state $\int\!d\omega
A_{coh}\!=\!Z_{\bf k}$ and $\int\!d\omega A_{incoh}\!=\!1\!-\!Z_{\bf
k}$ (where $A_{coh}$ and $A_{incoh}$ refer to coherent and
incoherent parts\index{Spectral function!coherent}\index{Spectral function!incoherent} of the spectral function), is:
\begin{equation}
\int_{-\infty}^{+\infty}d\omega A({\bf k},\omega)=1
\label{Comin:eq:intakw}
\end{equation}
\noindent
which reminds us that $A({\bf k},\omega)$ describes the
probability of removing/adding an electron with  momentum {\bf k}
and energy $\omega$ to a many-body system\index{Spectral function!sum rule}. However, as it also requires the knowledge of the electron addition part of the spectral function, it is not so useful in the analysis of ARPES
data, unless particle-hole symmetry holds. A sum rule\index{Spectral function!sum rule} more relevant to this task is:
\begin{equation}
\int_{-\infty}^{+\infty}d\omega f(\omega)A({\bf k},\omega)=n({\bf
k}) \label{Comin:eq:nk}
\end{equation}
\noindent
which solely relates the one-electron removal spectrum to the
momentum distribution $n({\bf k})$. When electronic correlations
are important and the occupation numbers are no longer good
quantum numbers, the discontinuity at ${\bf k}_F$ is reduced (as
discussed for the FL case) but a drop in $n({\bf k})$ is usually
still observable even for strong correlations \cite{Comin:nozieres:1}.
By tracking the {\it loci} of steepest descent of the
experimentally determined $n({\bf k})$ in $\mathbf{k}$-space, i.e., maxima in
$|\nabla\!_{\bf k}\,n({\bf k})|$, one may thus identify the Fermi
surface even in those correlated systems exhibiting particularly
complex ARPES features. However, great care is necessary in making
use of Eq.\,\ref{Comin:eq:nk} because the integral of Eq.\,\ref{Comin:eq:ikw}
does not give just $n({\bf k})$ but rather $I_0({\bf k},\nu,{\bf
A})n({\bf k})$ \cite{Comin:Damascelli:RMP,Comin:Damascelli:JESRP}.  A more detailed discussion of this point, with specific relevance to undoped Mott insulating cuprates, can be found in Ref.\,\citen{Comin:RonningF:PhoerF,Comin:ronning:1}.

\section{Matrix elements and finite resolution effects}
\label{Comin:sec:matele}
\index{ARPES!Matrix element}

As discussed in the previous section and summarized by
Eq.\,\ref{Comin:eq:ikw}, ARPES directly probes the one-particle spectral
function $A({\bf k},\omega)$. However, in extracting quantitative
information from the experiment, not only the effects of the matrix
element term $I_0({\bf k},\nu,{\bf A})$ have to be taken into
account, but also the finite experimental resolution and the
extrinsic continuous background due to the secondaries (those
electrons which escape from the solid after having suffered
inelastic scattering events and, therefore, with a reduced
$E_{kin}$). The latter two effects may be explicitly accounted for
by considering a more realistic expression for the photocurrent\index{ARPES!Cross-section} $I({\bf k},\omega)$:
\begin{equation}
\int\!\!d\tilde{\omega}d\tilde{{\bf k}}\,\left[ I_0(\tilde{{\bf
k}},\!\nu,\!{\bf A})f(\tilde{\omega})A(\tilde{{\bf
k}},\!\tilde{\omega})R(\omega\!-\!\tilde{\omega})Q({\bf
k}\!-\!\tilde{{\bf k}})\right]+B \label{Comin:eq:ikwreal}
\end{equation}
\noindent
which consists of the convolution of Eq.\,\ref{Comin:eq:ikw} with energy
($R$) and momentum ($Q$) resolution functions [$R$ is typically a
Gaussian, $Q$ may be more complicated], and of the background
correction B. Of the several possible forms for the background
function B \cite{Comin:hufner:1}, two are more frequently used: (i) the
step-edge background (with three parameters for height, energy
position, and width of the step-edge), which reproduces the
background observed all the way to $E_F$ in an unoccupied region
of momentum space; (ii) the Shirley background
$B_{Sh}(\omega)\!\propto\!\int^\mu_\omega d\omega^\prime
P(\omega^\prime)$, which allows to extract from the measured
photocurrent $I(\omega)\!=\!P(\omega)\!+\!c_{Sh} B_{Sh}(\omega)$
the contribution $P(\omega)$ of the unscattered electrons (with
only the parameter $c_{Sh}$ \cite{Comin:ShirleyDA:Hig-re}).
\begin{figure}[t!]
\centerline{\epsfig{figure=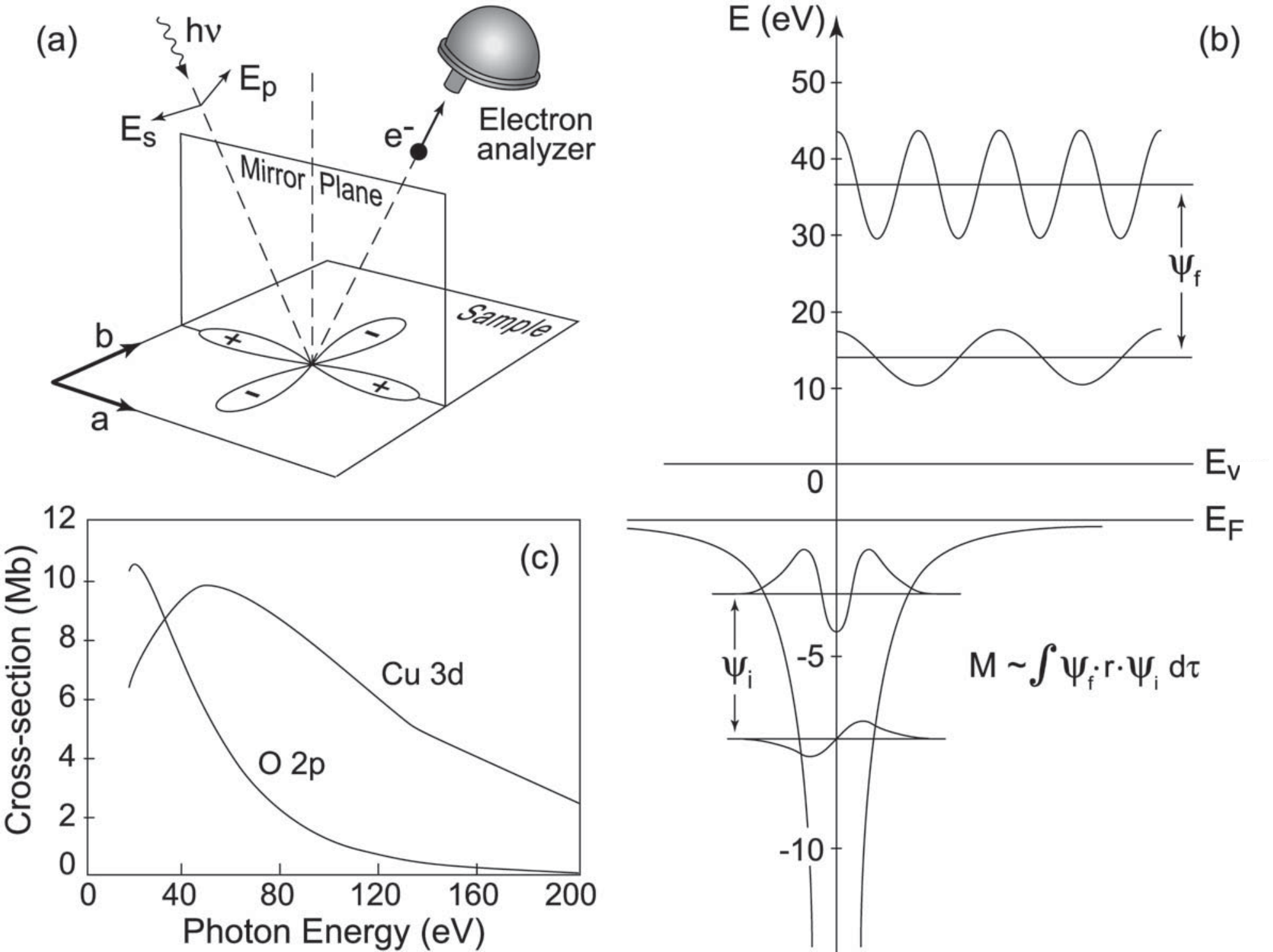,width=0.85\linewidth,clip=}}
\caption{(a) Mirror plane emission from a $d_{x^2-y^2}$ orbital.
(b) Sketch of the optical transition between atomic orbitals with
different angular momenta (the harmonic oscillator wavefunctions
are here used for simplicity) and free electron wavefunctions with
different kinetic energies (from Ref.\,\citen{Comin:Damascelli:PS}). (c)
Calculated photon energy dependence of the photoionization
cross-sections for Cu 3$d$ and O 2$p$ atomic levels.} \label{Comin:matrelem}
\end{figure}

Let us now very briefly illustrate the effect of the matrix
element term $I_0({\bf k},\nu,{\bf A})\!\propto\!|M_{f,i}^{\bf
k}|^2$, which is responsible for the dependence of the
photoemission data on photon energy and experimental geometry, and
may even result in complete suppression of the intensity
\cite{Comin:gobeligw:1,Comin:DietzE:Polda-,Comin:HermansonJ:Fin-st,Comin:EberhardtW:Dipsro}.
By using the commutation relation $\hbar{\bf p}/m\!=\!-i[{\bf
x},H]$, we can write $|M^{\bf k}_{f,i}|^2\!\propto\!|\langle
\phi_f^{\bf k}|\mbox{\boldmath$\varepsilon$}\!\cdot\!{\bf
x}|\phi_i^{\bf k}\rangle |^2$, where {\boldmath$\varepsilon$} is a
unit vector along the polarization direction of the vector
potential {\bf A}. As in Fig.\,\ref{Comin:matrelem}(a), let us consider
photoemission from a $d_{x^2-y^2}$ orbital, with the detector
located in the mirror plane (when the detector is out of the
mirror plane, the problem is more complicated because of the lack
of an overall well defined even/odd symmetry). In order to have
non vanishing photoemission intensity, the whole integrand in the
overlap integral must be an even function under reflection with
respect to the mirror plane. Because odd parity final states would
be zero everywhere on the mirror plane and therefore also at the
detector, the final state wavefunction $\phi_f^{\bf k}$ itself
must be even. In particular, at the detector the photoelectron is
described by an even parity plane-wave state $e^{i{\bf k \cdot r}}$ with
momentum in the mirror plane and fronts orthogonal to it
\cite{Comin:HermansonJ:Fin-st}. In turn, this implies that
$(\mbox{\boldmath$\varepsilon$}\!\cdot\!{\bf x})|\phi_i^{\bf
k}\rangle$ must be even. In the case depicted in
Fig.\,\ref{Comin:matrelem}(a), where $|\phi_i^{\bf k}\rangle$ is also even,
the photoemission process is symmetry allowed for {\bf A} even or
in-plane (i.e., $\mbox{\boldmath$\varepsilon$}_p\!\cdot\!{\bf x}$
depends only on in-plane coordinates and is therefore even under
reflection with respect to the plane) and forbidden for {\bf A}
odd or normal to the mirror plane (i.e.,
$\mbox{\boldmath$\varepsilon$}_s\!\cdot\!{\bf x}$ is odd as it
depends on normal-to-the-plane coordinates). For a generic initial
state of either even or odd symmetry with respect to the mirror
plane, the polarization conditions resulting in an overall even
matrix element\index{ARPES!Matrix element} can be summarized as:
\begin{equation}
\left\langle{\phi_f^{\bf k}}\right|{\bf A}\!\cdot\!{\bf
p}\left|{\phi_i^{\bf k}}\right\rangle \left\{{\matrix{{\phi_i^{\bf
k}\,\,even\,\,\left\langle + \right|+\left| + \right\rangle
\,\,\Rightarrow \,\,{\bf A}\,\,even} \vspace{0.1cm}\cr
{\!\!\!\phi_i^{\bf k}\,\,odd\,\,\,\,\,\left\langle +
\right|-\left| - \right\rangle \,\,\Rightarrow \,\,{\bf
A}\,\,odd}\cr }}\right.
\end{equation}

In order to discuss the photon energy dependence, from
Eq.\,\ref{Comin:eq:hint} and by considering a plane wave $e^{i{\bf kr}}$
for the photoelectron at the detector, one may more conveniently
write $|M^{\bf
k}_{f,i}|^2\!\propto\!|(\mbox{\boldmath$\varepsilon$}\!\cdot\!{\bf
k})\langle\phi_i^{\bf k}|e^{i{\bf kr}}\rangle |^2$.
The overlap integral, as sketched in Fig.\,\ref{Comin:matrelem}(b),
strongly depends on the details of the initial state wavefunction
(peak position of the radial part and its oscillating character), and on the wavelength of the outgoing plane wave. Upon increasing the photon energy, both $E_{kin}$ and {\bf k} increase, and $M^{\bf k}_{f,i}$ changes in a fashion which is not necessarily monotonic (see Fig.\,\ref{Comin:matrelem}(c), for the Cu 3$d$ and the O 2$p$ atomic case). In fact, the photoionization cross section is
usually characterized by one minimum in free atoms, the so-called
Cooper minimum \cite{Comin:Cooperjw:1}, and a series of them in solids
\cite{Comin:MolodtsovSL:Coomtp}.

\section{State-of-the-art photoemission} \label{Comin:sec:stateart}
\index{ARPES!the technique}

\begin{figure}[t!]
\centerline{\epsfig{figure=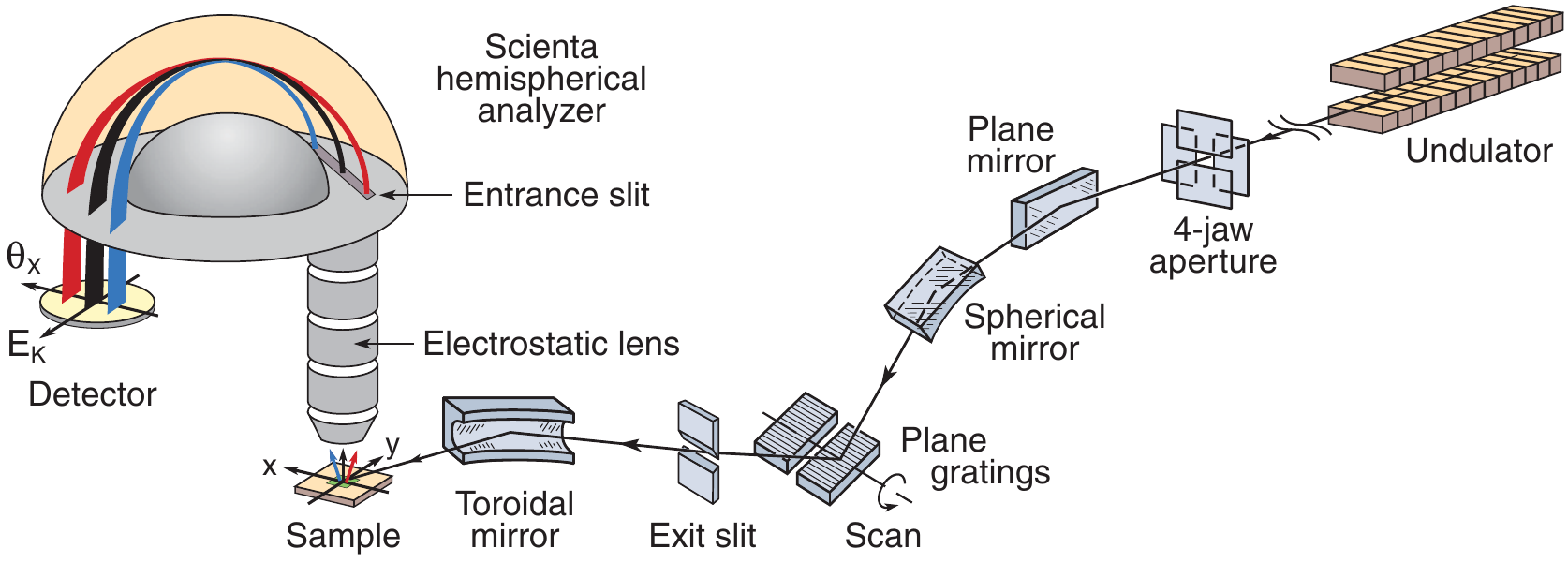,width=1\linewidth,clip=}}
\caption{(Color). Beamline equipped with a plane grating
monochromator and a 2D position-sensitive electron analyzer (from Ref.\,\citen{Comin:Damascelli:RMP}).}
\label{Comin:scienta}
\end{figure}

The configuration of a generic angle-resolved photoemission
beamline is shown in Fig.\,\ref{Comin:scienta}.
A beam of radiation peaked about a specific photon energy is produced in a wiggler or an undulator
(these so-called `insertion devices' are the straight sections of
the electron storage ring where radiation is produced). The light
is then monochromatized at the desired photon energy by a grating
monochromator, and focused on the sample. Alternatively, a UV-laser or a gas-discharge lamp can be used as a radiation source (the latter has to be
properly monochromatized, to avoid complications due to the
presence of different satellites, and refocused to a small spot
size, essential for high angular resolution). However, synchrotron
radiation offers important advantages: it covers a wide spectral
range (from the visible to the X-ray region) with an intense and
highly polarized continuous spectrum; lasers and discharge lamps
provide only a few resonance lines at discrete energies.

Photoemitted electrons are then collected by the electron analyzer, where
kinetic energy and emission angle are determined (the whole system
is in ultra-high vacuum at pressures lower than
5$\times\!10^{-11}$ torr).
A conventional hemispherical analyzer\index{ARPES!Electron analyzer} consists of a multi-element electrostatic input lens, a hemispherical deflector with entrance and exit slits, and an electron detector (i.e., a channeltron or a
multi-channel detector). The heart of the analyzer is the
deflector which consists of two concentric hemispheres (of radius
$R_1$ and $R_2$). These are kept at a potential difference $\Delta
V$, so that only those electrons reaching the entrance slit with
kinetic energy within a narrow range centered at
$E_{pass}\!=\!e\Delta V/(R_1/R_2\!-\!R_2/R_1)$ will pass through
this hemispherical capacitor, thus reaching the exit slit and then
the detector. This way it is possible to measure the kinetic
energy of the photoelectrons with an energy resolution given by
$\Delta E_a\!=\!E_{pass}(w/R_0\!+\!\alpha^2\!/4)$, where
$R_0\!=\!(R_1\!+\!R_2)/2$, $w$ is the width of the entrance slit,
and $\alpha$ is the acceptance angle. The role of the
electrostatic lens is that of decelerating and focusing the
photoelectrons onto the entrance slit. By scanning the lens
retarding potential one can effectively record the photoemission
intensity versus the photoelectron kinetic energy. 
\begin{figure}[t!]
%\centerline{\epsfig{file=Sr2RuO4_3D_Christian.pdf,bbllx=0,bblly=0,bburx=300,bbury=625,clip=,angle=0,width=1\linewidth}}
\begin{center}
\includegraphics*[trim=0 0 0 0,clip,width=0.9\textwidth]{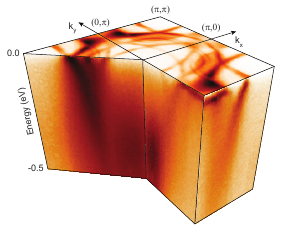}
\caption{(Color). 3D plot of the ARPES intensity from
Sr$_2$RuO$_4$ versus energy ($\omega$) and momentum (${k}_{x}$ and ${k}_{y}$), in the first Brillouin zone. The top plane shows the Fermi surface $ A (\mathbf{k}, \omega = 0) $, while the side cuts show the spectral function along certain high-symmetry directions. The multiple features that are visible in dispersion and Fermi surface, in addition to the three Ru-${t}_{2g}$ bands expected for this material, arise from near-surface structural instabilities, and associated structural reconstruction and band-folding \cite{Comin:Veenstra:Sr2RuO4}. Note that these data differ from those in Fig.\,\ref{Comin:4d_oxides}, which were measured under different conditions intended to suppress the contribution from the surface top-most layers and are thus representative of the bulk of Sr$_2$RuO$_4$ \cite{Comin:DamascelliA:Fersss}.} %This plot showcases the capabilities of state-of-the-art high-resolution spectrometers (adapted from Ref.\,\citen{Comin:Veenstra:Sr2RuO4}).}
\label{Comin:Sr2RuO4_3D}
\index{ARPES!Electron analyzer}
\end{center}
\end{figure}

One of the innovative characteristics of a state-of-the-art analyzer is the two-dimensional position-sensitive detector consisting of two
micro-channel plates and a phosphor plate in series, followed by a
CCD camera. In this case, no exit slit is required: the electrons,
spread apart along the $Y$ axis of the detector
(Fig.\,\ref{Comin:scienta}) as a function of their kinetic energy due to
the travel through the hemispherical capacitor, are detected
simultaneously [in other words, a range of electron energies is
dispersed over one dimension of the detector and can be measured
in parallel; scanning the lens voltage is in principle no longer
necessary, at least for narrow energy windows (a few percent of
$E_{pass}$)]. Furthermore, contrary to a conventional electron
spectrometer in which the momentum information is averaged over
all the photoelectrons within the acceptance angle (typically
$\pm1^\circ$), state-of-the-art 2D position-sensitive electron
analyzers can be operated in angle-resolved mode, which provides
energy-momentum information not only at a single $k$-point but
along an extended cut in $k$-space. In particular, the photoelectrons within a variable angular window as wide as $\sim\!30^\circ$ along the direction defined by the analyzer entrance slit can be focused on different $X$ positions on the
detector (Fig.\,\ref{Comin:scienta}). It is thus possible to measure multiple energy distribution curves simultaneously for different photoelectron angles, obtaining a 2D snapshot of energy versus momentum. Such snapshots, for differing sample orientations, can be combined to form a 3D volume (Fig.\,\ref{Comin:Sr2RuO4_3D}), and then cut at constant energy to generate a material's Fermi surface when done at $\omega\!=\!{E}_{F}$ on a metal, or along any $k$-space path to generate a band mapping versus energy and momentum. 
\begin{figure}[t!]
\centerline{\epsfig{figure=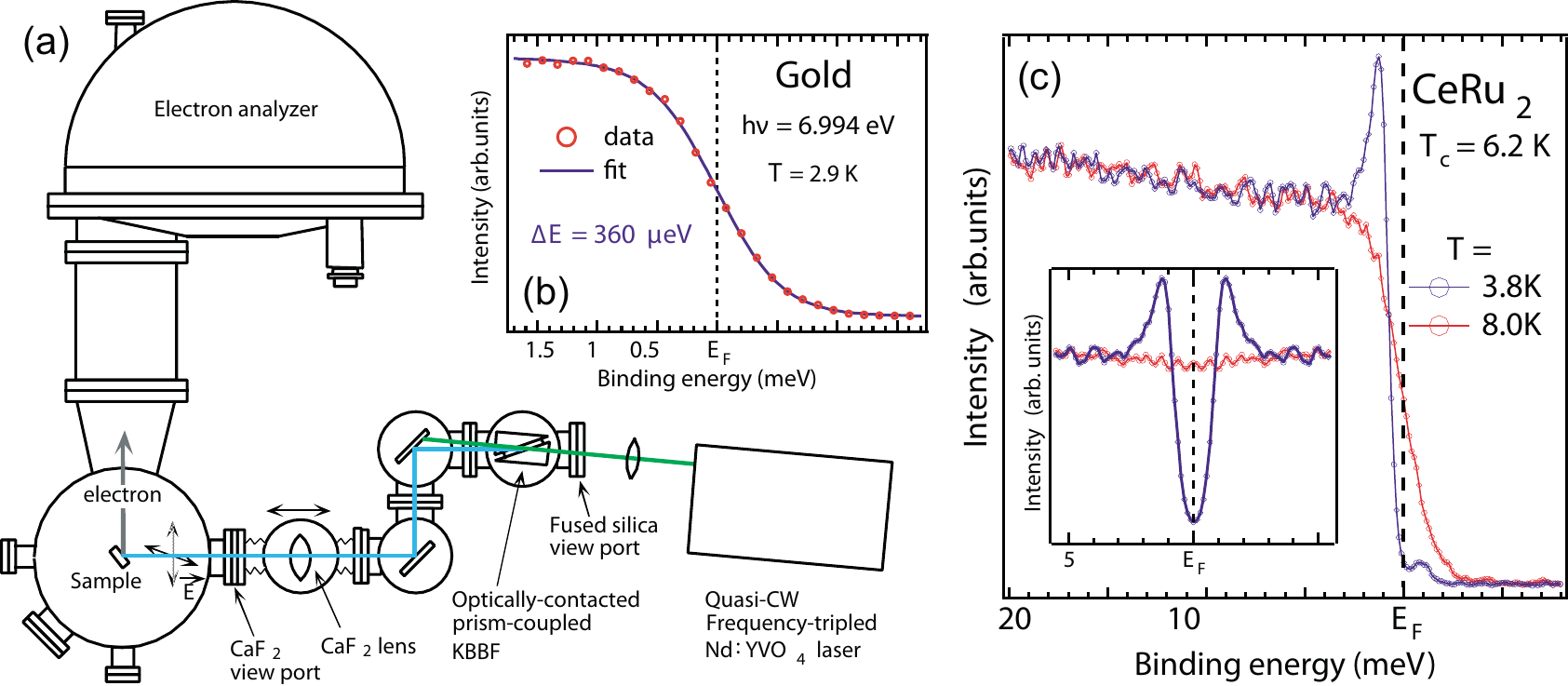,width=1\linewidth,clip=}}
\caption{(Color). (a) Schematics of a laser-ARPES setup, with (b) sub-meV resolution on a Au target ($\Delta E\!=\!360\,\mu $eV). This enables (c) the detection of a $\Delta\!=\!500\,\mu$eV superconducting gap in the heavy fermion system CeRu${}_{2}$ (from Ref.\,\citen{Comin:Kiss:CeRu2}).}
\label{Comin:scienta_hires}
\end{figure}
State-of-the-art spectrometers typically allow for energy and angular resolutions of less than approximately 1 meV and $0.1-0.2^\circ$, respectively\index{ARPES!Experimental resolution}. Taking as example the transition metal oxides and in particular the cuprate superconductors (for which $2\pi/a\!\simeq\!1.6$ \AA$^{-1}$), one can see from
Eq.\,\ref{Comin:eq:momentum} that 0.2$^\circ$ corresponds to $\sim$0.5\% of the Brillouin zone size, for the 21.2 eV photons of the He-I$\alpha$ line typically used in ARPES systems equipped with a gas-discharge lamp. In the case of a beamline, to estimate the total energy resolution one has to take into account also the $\Delta E_{m}$ of the
monochromator, which can be adjusted with entrance and exit slits (the ultimate resolution a monochromator can deliver is given by its resolving power $R\!=\!E/\Delta E_{m}$ and in general worsens upon increasing the photon energy). The current record in energy resolution is of 360 $\mathrm{\mu}$eV obtained on an ARPES spectrometer equipped with a Scienta R4000 electron analyzer and a UV laser operating in continuous-wave mode at $\sim\!6.994\,$eV (see Fig.\,\ref{Comin:scienta_hires}). One should note however that while the utilization of UV lasers allows superior resolutions, it also leads to some important shortcomings: (i) access to a limited region of k-space, often smaller than the first Brillouin zone, due to the reduced kinetic energy of photoelectrons; (ii) extreme sensitivity to final state effects, with the detailed energy-momentum structure of the final states becoming important, since at low photon energies one cannot reach the high-energy continuum (the kinematic constraints of energy and momentum conservation may be satisfied only for a limited set of momenta, and as a result  the photoemission intensity might be completely suppressed in certain regions of the Brillouin zone); (iii) breakdown of the sudden approximation, in which case the photoemission intensity would all be found in the ``0-0" transition between the initial and final ground states (see discussion of Fig.\,\ref{Comin:geometry} and especially \ref{Comin:H2_model}), providing no information on the excited states of the system left behind, and in turn on the strength and nature of the underlying many-body interactions (the crossover from sudden to adiabatic regime in TMOs is still being debated, see e.g. Ref.\,\citen{Comin:Koralek:adiabatic}, and depends on the specific relaxation processes of a given material).

\section{Physics of correlations - the ARPES perspective} \label{Comin:sec:correlations}

The sensitivity of ARPES to correlation effects is deeply connected to its corresponding observable, which is the one-particle spectral function previously introduced in Sec.\,\ref{Comin:sec:spectral}. This physical quantity conveys information not only on the single-particle excitations, but also on the many-body final states which can be reached in the photoemission process. However, the distinction between single-particle and many-body features in $ A (\mathbf{k}, \omega) $ at the experimental level is often subtle. In order to disentangle the nature of the underlying excitations it is common practice to decompose the spectral function into a coherent, $ {A}_{coh} (\mathbf{k}, \omega) $, and an incoherent part, $ {A}_{incoh} (\mathbf{k}, \omega) $, as explained in Sec.\,\ref{Comin:sec:spectral}.\index{Correlations} 

Whereas in a purely non-interacting system all single-electron excitations are coherent, since they are insensitive to the behavior of the other particles, things can be quite different when \textit{electron-electron} interactions, and therefore quantum-mechanical correlation effects, are turned on. For these reasons, the redistribution of spectral weight between the coherent and incoherent part in $ A (\mathbf{k}, \omega) $ is commonly regarded as a distinct signature of correlations at work. In particular, the integrated spectral intensity of the coherent part, the \textit{quasiparticle strength} $Z_{\bf
k}$ that can be extracted from ARPES, is a relatively direct measure of the correlated behavior of a given system. Following its definition given in Eq.\,\ref{Comin:eq:gwk}, $Z_{\bf
k}$ can vary from 1 (non-interacting case) to 0 (strongly correlated case, no coherent states can be excited).

In the following subsections we will explain how the concept of correlation 
already emerges in simple molecular-like systems (i.e., few-body) and evolves 
into the complex structures found in solid-state materials (i.e., many-body). Different types of correlation effects will be reviewed, with particular emphasis on those stemming from \textit{electron-phonon} and \textit{electron-electron} interactions. We will then discuss different aspects of correlated electron behavior in a few selected transition metal oxides, and show how correlations evolve with -- and to some degree can be controlled by -- the external control parameters introduced in Sec.\,\ref{Comin:sec:intro_correlations} and Fig.\,\ref{Comin:Periodic_table}.

\subsection{Origin of correlations in photoelectron spectroscopy}

%Introduce here the idea of correlations in photoelectron spectra using analogy to Franck-Condon principle (refer to Fig.\,\ref{Comin:H2_model})

In general, the connection between the one-particle spectral function and correlations is not immediately obvious and might look mysterious to the reader. It is useful and instructive to clarify what the photoelectron spectrum for a correlated system looks like, beginning with an example from molecular physics.
\begin{figure}[!t]
\centerline{\epsfig{figure=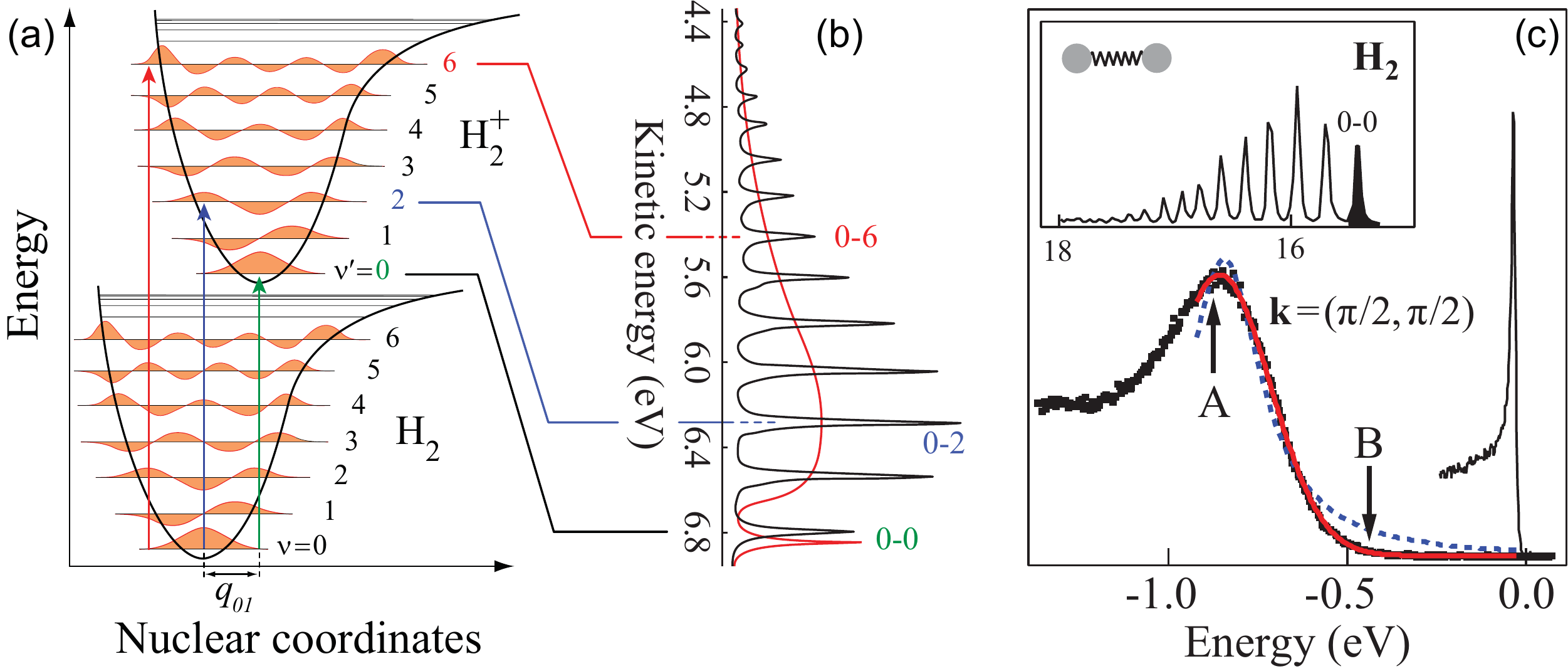,width=1\linewidth,clip=}}
\caption{(Color). (a) Franck-Condon effect and its relation to the single-particle spectral function in atomic physics. (b) Photoionization spectrum of gaseous H${}_{2}$ (from Ref.\,\citen{Comin:Asbrink:H2}), featuring a comb of lines corresponding to the various accessible excited (vibrational) final states of the H${}_{2}^{+}$ system left behind; the red line is an abstraction to what would happen in solid-state H${}_{2}$ (from Ref.\,\citen{Comin:sawatzky:1}). (c) $ A (\mathbf{k}\!=\!(\pi/2,\pi/2),\omega) $ from Ca${}_{2}$CuO${}_{2}$Cl${}_{2}$, showing the broad incoherent Zhang-Rice peak, with the sharp Sr${}_{2}$RuO${}_{4}$ lineshape superimposed for comparison (from Ref.\,\citen{Comin:Kyle:polaronCCOC}).}
\index{Franck-Condon effect}\index{Polaron}
\label{Comin:H2_model}
\end{figure}
In Fig.\,\ref{Comin:H2_model} we show the photoionization spectrum of the molecular gas H${}_{2}$ which, at variance with a simpler atomic gas (e.g., He or Ne), exhibits a fine structure made of a series of peaks almost evenly separated in energy. The underlying physical explanation for these spectral features relates to the Franck-Condon principle, which is explained in Fig.\,\ref{Comin:H2_model}(a). This is best understood if we write down the equation for the photoionization cross section, which will involve: (i) an initial state wavefunction ${\psi}_{i}$, assumed to be the ground state for the neutral molecule; and (ii) a final state wavefunction ${\psi}_{f}$, which can be a linear combination of eigenstates for the ionized, positively charged molecule H${}_{2}^{+}$. The possible eigenstates we consider here can be separated into an electronic part (the hydrogen-like 1s orbital $ {\phi}_{1s} $) and a nuclear part, which in a diatomic molecule like H${}_{2}$ can be vibrationally excited. The latter is given, to a good approximation, by one of the eigenfunctions of the harmonic oscillator, $ {\phi}_{n} ({R}_{eq}) $, which depend on the interatomic equilibrium distance ${R}_{eq}$. Combining together electronic and nuclear (i.e., vibrational) components we obtain a basis set for the molecular Hamiltonian in the form $ {\psi}_{n}\!=\!{\phi}_{1s} {\phi}_{n} ({R}_{eq}) $. We can then use this set of functions in the matrix element governing the photoionization process\index{Franck-Condon effect!${\mathrm{H}}_{2}$ molecule}:

\begin{equation}
{I}_{{\mathrm{H}}_{2} \rightarrow {\mathrm{H}}_{2}^{+}} \propto \sum_{m} \langle {\psi}_{m}^{{\mathrm{H}}_{2}^{+}} \vert \mathbf{p} \cdot \mathbf{A} \vert {\psi}_{n=0}^{{\mathrm{H}}_{2}} \rangle \propto \sum_{m} \langle {\phi}_{\mathbf{k}, m} \vert \mathbf{p} \cdot \mathbf{A} \vert {\phi}_{1s} \rangle \langle {\phi}_{m} ({R}_{eq}^{{\mathrm{H}}_{2}^{+}}) \vert {\phi}_{0} ({R}_{eq}^{{\mathrm{H}}_{2}}) \rangle\,.
\label{Comin:eq:H2_photoionization}
\end{equation}
\noindent
Here $ \mathbf{p} \cdot \mathbf{A} $ is the dipole interaction operator and the initial state is the ground state for H${}_{2}$, or  $ {\psi}_{GS}^{{\mathrm{H}}_{2}}\!=\!{\psi}_{n=0}$. The term $\langle {\phi}_{\mathbf{k},m} \vert \mathbf{p} \cdot \mathbf{A} \vert {\phi}_{1s} \rangle\!=\!{M}_{m}^{\mathbf{k}}$ is the \textit{matrix element} previously introduced in Sec.\,\ref{Comin:sec:matele}, representing the overlap between the initial-state electronic wavefunction $ {\phi}_{1s} $ and the final state plane-wave $ {\phi}_{\mathbf{k}, m} $. It is readily seen that, if $ {R}_{eq}^{{\mathrm{H}}_{2}^{+}}\!=\!{R}_{eq}^{{\mathrm{H}}_{2}} $, then $ {I}_{{\mathrm{H}}_{2} \rightarrow {\mathrm{H}}_{2}^{+}}\!\propto\!\sum_{m} {\delta}_{m,0} $ and the photoionization spectrum would be composed of a single peak, corresponding to the ``0-0" transition between the initial and final ground states. In reality, the neutral and ionized molecule will see a different charge distribution (thus leading to a different electrostatic potential), due to the missing Coulomb interaction term for the 1\textit{s} electrons in the Hamiltonian for H${}_{2}^{+}$. As a consequence, the molecule before and after photoexcitation will have a different interatomic equilibrium distance, and many of the terms in Eq.\,\ref{Comin:eq:H2_photoionization} will be different from zero resulting in multiple transitions in the experimental spectrum [corresponding to the vertical excitations in Fig.\,\ref{Comin:H2_model}(a) and to the various peaks in Fig.\,\ref{Comin:H2_model}(b)]. The lowest energy peak [labeled ``0-0" in Fig.\,\ref{Comin:H2_model}(b)] still corresponds to a transition into the ground state of the ionized molecule, but it only contains a fraction of the total photoemission intensity, or \textit{spectral weight}. At this point it is useful to introduce an alternative definition (but equivalent to the one given in Sec.\,\ref{Comin:sec:spectral}) of \textit{coherent} and \textit{incoherent} spectral weight\index{Spectral weight!coherent}\index{Spectral weight!incoherent}: 

\begin{itemize}
\item The \textit{coherent} spectral weight is a measure of the probability to reach the \textit{ground state} of the final-state Hamiltonian ($ {\mathcal{H}}_{{\mathrm{H}}_{2}^{+}} $) in the photoexcitation process. In experimental terms, it is represented by the total area of the 0-0 transition shown in Fig.\,\ref{Comin:H2_model}(b).
\item The \textit{incoherent} spectral weight is a measure the probability to leave the ionized system in any of its \textit{excited states}. It can be therefore calculated from the integrated intensity of all the 0-\textit{m} ($ m \neq 0 $) peaks in the ionization spectrum, as shown in Fig.\,\ref{Comin:H2_model}(b).
\end{itemize}

In solid-state, many-body systems, both molecular vibrations and electronic levels are no longer discrete but have an energy dispersion (turning into phonons and electronic bands, respectively). This is what gives a continuum of excitations when many body interactions are at play, as opposed to the sharp excitation lines of the ${\mathrm{H}}_{2}$ case. However these concepts remain valid, although we shall now restate them within a many-body framework\index{Quasiparticle weight}: 

\begin{itemize}
\item The \textit{coherent} spectral weight corresponds to the probability of reaching, via the electron addition/removal process ($ \Delta N\!=\!\pm 1 $, where \textit{N} is the initial number of electrons), the \textit{many-body} ground-state for the $ (N \pm 1) $-particle Hamiltonian ($ {\mathcal{H}}_{N\pm1}^{\mp} $).\index{Spectral weight!coherent}
\item The \textit{incoherent} spectral weight gives the cumulative probability that the $(N \pm 1)$-particle system is instead left in an excited state.\index{Spectral weight!incoherent}
\end{itemize}

%Here make the connection to solid-state: electron-nucleus interaction in H${}_{2}$ becomes el-ph coupling, 0-0 transition become QP pole, discrete spectrum becomes continuum of excitations (again refer to Fig.\,\ref{Comin:H2_model}).
%Building on this analogy introduce concepts of \textit{coherent} and \textit{incoherent} spectral weight.\\

A photoelectron spectrum for a solid-state many-body system will look like the dashed curve in Fig.\,\ref{Comin:H2_model}(b), due to the multitude of final states that can be reached as a result of the photoemission process. The well-defined features characterizing $ A (\mathbf{k}, \omega) $ in the molecular case will then broaden out into a continuum of excitations. This was experimentally found to occur in the strongly-coupled cuprate material Ca${}_{2}$CuO${}_{2}$Cl${}_{2}$ [see Fig.\,\ref{Comin:H2_model}(c)], which will be discussed in more detail in the next section. We also note that for the incoherent part of the spectral function two cases are possible in a solid:

\begin{enumerate}
\item $ {A}_{incoh} (\vec{k}, \omega) $ is composed of gapless many-body excitations, e.g. creation of electron-hole pairs in a metal; this typically produces an asymmetric lineshape, as in the case of the Doniach-Sunjic model \cite{Comin:Doniach:lineshape}.\index{Spectral function!Doniach-Sunjic}
\item $ {A}_{incoh} (\vec{k}, \omega) $ originates from gapped excitations, e.g. coupling between electrons and optical phonons; in this case the coherent part is well separated from the incoherent tail, and a quasiparticle peak can be more properly identified.
\end{enumerate}

What we have just seen for the $ {\mathrm{H}}_{2} $ molecule stems from the interaction between the electronic and the nuclear degrees of freedom. In the absence of such interplay there would be no fine structure in the corresponding spectral function. This is a very important concept, which is deeply connected to the idea of correlations. The Hamiltonian of a given (few-body or many-body) system, in the absence of interaction terms, can be decomposed into a sum of single particle terms (non-interacting case). Correspondingly, the system is unperturbed by the addition or removal of a particle during the photoexcitation process; due to the orthonormality of the involved eigenstates, the $ (N \pm 1) $-system left behind will not be found in a superposition of excited states but rather left unperturbed in its ground state (at zero temperature). Single-particle spectroscopy would then detect a single transition (e.g., the 0-0 peak in Fig.\,\ref{Comin:H2_model}) and the spectral weight is \textit{fully coherent}.
Conversely, when the electron-nucleus and/or electron-lattice interaction are switched on, addition/removal of a single electron perturbs the molecular/lattice potential to some degree and this can trigger creation or annihilation of one or multiple vibrational modes in the process. As we will see in the following section for electron-phonon coupling in solids, the effect at the level of spectral function can be very different according to the strength of the interaction.

\subsection{Electron-phonon correlations in solids: the polaron}

The interaction of the mobile charges with the static ionic lattice is what underlies the formation of electronic bands in all crystalline materials. However, the lattice is never really static and its low-energy excitations, the \textit{phonons}, are present even at very low temperatures. As they hop around in the lattice, electrons can interact (through the ionic Coulomb potential) with -- or become ``dressed" by -- phonons, thereby slowing their quantum motion. These new composite entities, known as \textit{polarons}, represent the true quasiparticles of the coupled electron-lattice system: the properties of the ``bare" electrons, \textit{in primis} bandwidth and mass, are now renormalized in a fashion which directly depends on the strength of the \textit{electron-phonon} coupling. Here we show two examples of polaronic physics, one experimental and the other theoretical, which exhibit different features but relate to the same underlying \mbox{interactions}\index{Transition Metal Oxides!3\textit{d}-based}\index{Polaron}.\index{Electron-phonon coupling}\\
The first case is that of Ca${}_{2}$CuO${}_{2}$Cl${}_{2}$ (CCOC)\index{Cuprates!Ca${}_{2}$CuO${}_{2}$Cl${}_{2}$}. This compound is the undoped \textit{parent} compound of the high-${T}_{c}$ superconducting cuprates, where the low-energy physics originates from the hybridized Cu 3\textit{d}\,--\,O 2\textit{p} states of the CuO${}_{2}$ planes. In Fig.\,\ref{Comin:H2_model}(c), the ARPES spectrum of CCOC at $ \vec{\mathrm{k}}\!=\!(\pi/2, \pi/2) $ is shown \cite{Comin:Kyle:polaronCCOC}. This value of electron momentum corresponds to the lowest ionization state of the Zhang-Rice singlet (ZRS) band \cite{Comin:ZhangFC:EffHsC}\index{Zhang-Rice singlet}. The latter is a 2-particle state, made of a combination of one O 2\textit{p} and one Cu 3\textit{d} hole in a total spin-zero state (singlet). Hence the nature of such a state is intrinsically correlated, and cannot be described within a single-particle framework. While this feature was originally recognized as the quasiparticle pole of the same Cu--O band that is also found in the doped compounds \cite{Comin:WellsBO:Evkrmb}, the Gaussian lineshape, together with the broad linewidth ($ \Gamma\!\sim\!0.5 $\,eV) suggest that this feature might instead be identified as the incoherent part of the spectral function. It follows that the spectral function has no actual quasiparticle weight $ Z_{\bf
k} $, because the intensity of the lowest energy excitation [the ``0-0" line marked by the ``B" arrow in Fig.\,\ref{Comin:H2_model}(c)] approaches zero\index{Quasiparticle weight}.
\begin{figure}[!t]
\centerline{\epsfig{figure=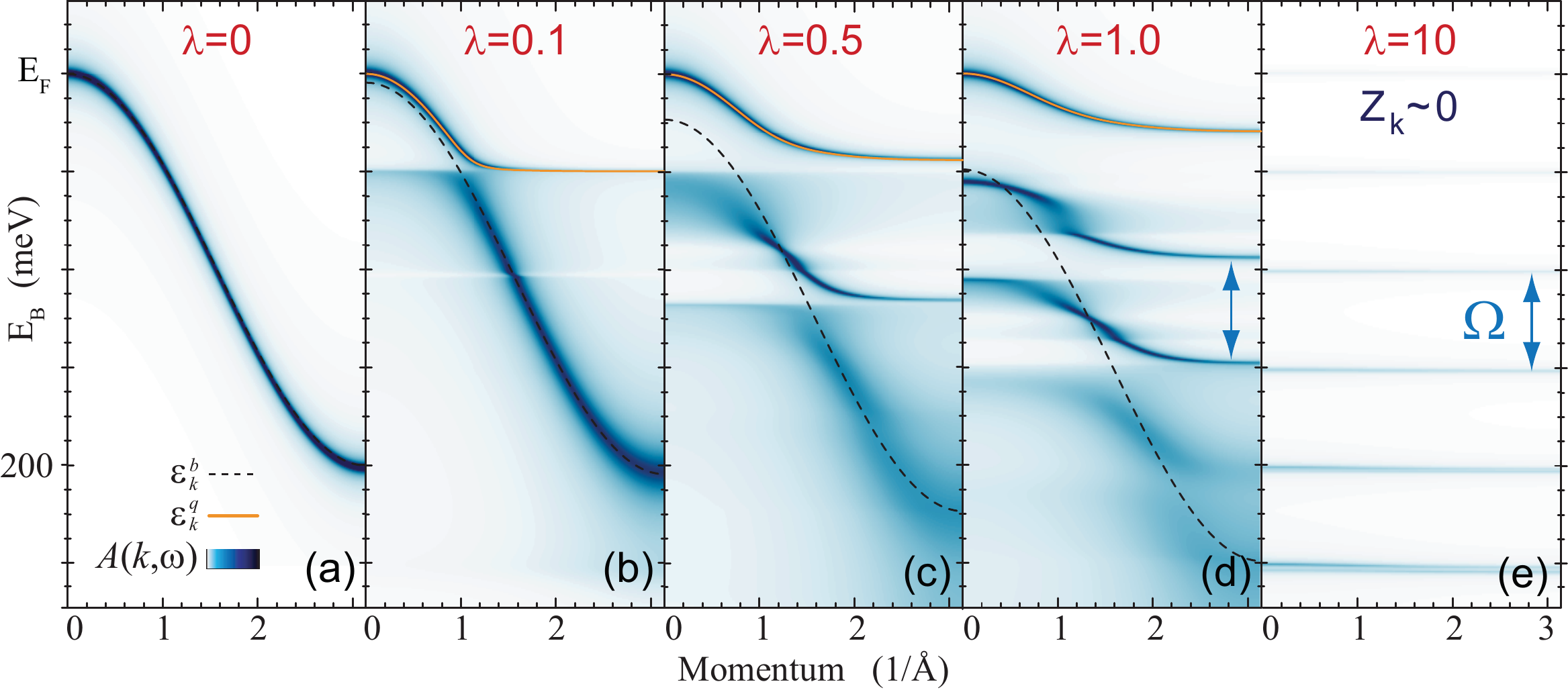,width=1\linewidth,clip=}}
\caption{(Color). $ A (\mathbf{k},\omega) $ for the Holstein model, showing the quasiparticle band and the 0-, 1- and \textit{n}-phonon dispersing poles in the spectral function for increasing values of the electron-phonon coupling $\lambda\!=\,$0, 0.1, 0.5, 1.0 and 10 (panels a, b, c, d and e, respectively). The dashed black, and the orange lines represent the bare band ${\varepsilon}_{\mathrm{k}}^{b}$, and the renormalized quasiparticle band ${\varepsilon}_{\mathrm{k}}^{q}$, respectively. The Fermi energy ${E}_{F}$ has been set at the top of the quasiparticle band (from Ref.\,\citen{Comin:VeenstraCN:SF_Holstein}).}
\index{Zhang-Rice singlet}
\label{Comin:Polaron}
\end{figure}

While the previous example illustrates a case where the strong electron-boson interaction entirely washes away the coherent spectral weight, in different physical systems it is possible to have sizeable weight in the quasiparticle pole. This is illustrated with the second example discussed here, the one-dimensional (1D) Holstein model \cite{Comin:VeenstraCN:SF_Holstein}, shown in Fig.\,\ref{Comin:Polaron}. This model is represented by the following Hamiltonian:

\begin{equation}
{\mathcal{H}}_{\mathrm{Holstein}}^{1D} = {\sum}_{\mathrm{k}} {\varepsilon}_{\mathrm{k}}^{b} {c}_{\mathrm{k}}^{\dagger} {c}_{\mathrm{k}} + \Omega {\sum}_{\mathrm{Q}} {b}_{\mathrm{Q}}^{\dagger} {b}_{\mathrm{Q}} + \frac{g}{\sqrt{n}} {\sum}_{\mathrm{k},\mathrm{Q}} {c}_{\mathrm{k}-\mathrm{Q}}^{\dagger} {c}_{\mathrm{k}} ( {b}_{\mathrm{Q}}^{\dagger} + {b}_{-\mathrm{Q}} )\,.
\label{Comin:eq:Holstein}
\end{equation}
\noindent
The evolution of the associated spectral function as a function of the dimensionless electron-phonon coupling parameter $ \lambda\!=\!{g}^{2} / 2 t \Omega $ (in this calculation $ \Omega\!=\!50$\,meV) is shown in Fig.\,\ref{Comin:Polaron}. The case $ \lambda\!=\!0 $ (no coupling), shown in Fig.\,\ref{Comin:Polaron}(a), yields a spectral function which exactly follows the bare electronic band $ {\varepsilon}_{\mathrm{k}}^{b} $ (indicated by the black dashed line in Fig.\,\ref{Comin:Polaron}), i.e. $ A (\mathrm{k}, \omega)\!=\!\delta (\omega - {\varepsilon}_{\mathrm{k}}^{b}) $, where the $\delta$-function is here broadened into a Lorentzian for numerical purposes. Already in the small coupling limit $ \lambda\!=\!0.1 $ a quasiparticle band branches off the original band, with a k-dependent spectral weight [see Fig.\,\ref{Comin:Polaron}(b)]. The latter is substantially redistributed, with the spectral weight at binding energies higher than the quasiparticle band belonging to the incoherent part of the spectral function, which forms a continuum of many-body excitations for $ {E}_{B}\!>\!\Omega $. As the electron-phonon coupling is further increased, it can be noted from Fig.\,\ref{Comin:Polaron}(c,d) how: (i) there is a progressive renormalization of the quasiparticle band ($ {\varepsilon}_{\mathrm{k}}^{q} $), which implies a reduction in the total bandwidth and a change in the slope $ \partial {\varepsilon}_{\mathrm{k}}^{q} / \partial \mathrm{k} $ (quasiparticle velocity); (ii) the spectral weight is redistributed between the quasiparticle band $ {\varepsilon}_{\mathrm{k}}^{q} $ and the incoherent features at higher binding energy, corresponding to a photohole copropagating with, or ``dressed" by, one or multiple phonons. In the strong-coupling regime $ \lambda\!=\!10 $, the quasiparticle band and its \textit{n}-phonon replicas are nondispersive (i.e., corresponding to a diverging quasiparticle mass), and the coherent spectral weight $ Z_{\bf
k} $ has almost completely vanished [Fig.\,\ref{Comin:Polaron}(d)].

After having discussed quasiparticle renormalization due to electron-phonon coupling, in the following we will turn our focus onto electron-electron interaction 
effects, which are particularly pronounced in 3\textit{d}-TMOs, and dominate the low-energy electrodynamics in these systems.

\subsection{Doping-controlled coherence: the cuprates}
\index{Spectral weight!coherent!vs. doping}
\index{Transition Metal Oxides!3\textit{d}-based}

%At this point talk about the use of doping to control SWT. Describe the case of cuprates (YBCO) and how coherence in the spectral weight evolve in the phase diagram. Emphasize the K-evaporation technique as an efficient method to dope a material in situ and control its electronic structure without altering its other properties (lattice parameters, symmetry, stcohiometry, etc.). Refer to Fig.\,\ref{Comin:YBCO1} and Fig.\,\ref{Comin:YBCO2} (although maybe I can incorporate them in a single figure, and I guess I would keep the story of $Z_{\bf k}$ measured by EDC area and only marginally mention the use of bilayer splitting, which in any case is limited only to YBCO)
%\begin{figure}[!htbp]
%\centerline{\epsfig{figure=YBCO_david_1.pdf,width=1\linewidth,clip=}}
%\caption{(Color). Spectral weight transfer signatures in one-particle spectroscopy: ARPES on YBCO.}
%\label{Comin:YBCO1}
%\end{figure}

As anticipated, copper-based oxide superconductors exhibit a rich phase diagram, encompassing a variety of unconventional phases (Fig.\,\ref{Comin:Periodic_table}), which include: high-temperature superconductivity, Mott insulating behavior, pseudogap phase, strange metal (non-conventional Fermi liquid), and possibly electronic liquid crystal (nematic phase), to name a few. In particular, their remarkable peculiarity lies in the possibility of realizing these different phases simply by controlling the charge carriers doped into the CuO${}_{2}$ planes.

A manifestation of the underlying correlated nature of these materials can be found in the doping-dependent evolution of coherent behavior in the low-energy electrodynamics. This is the case of the ARPES results on YBa${}_{2}$Cu${}_{3}$O${}_{6+x}$ (YBCO),\index{Cuprates!YBa${}_{2}$Cu${}_{3}$O${}_{6+x}$} one of the most studied within the family of cuprates owing to its superior purity. In this and similar materials, hole-doping is usually controlled at the chemical level, by tuning the stoichiometric ratio between the O and Cu content. As for the study of the low-energy electronic structure by ARPES, this has been hampered by the lack of a natural cleavage plan and especially the polarity of the material, which leads to the the self doping of the cleaved surfaces \cite{Comin:Borisenko:YBCO,Comin:Hossain:YBCO,Comin:David:YBCO}. As a result, while the bulk of YBCO cannot be doped beyond 20\% by varying the oxygen content, the surfaces appear to be overdoped up to almost 40\% (the highest overdoping value reached on any cuprate \cite{Comin:David:YBCO}). An approach devised to resolve this problem involves the control of the carrier concentration at the surface \cite{Comin:Hossain:YBCO,Comin:David:YBCO} by in-situ potassium deposition on the cleaved crystals, which enables the investigation of the surface electronic structure all the way from the overdoped ($p\!\sim\!0.37$) to the very underdoped region of the phase diagram ($p\!\sim\!0.02$).\\
\begin{figure}[!t]
\centerline{\epsfig{figure=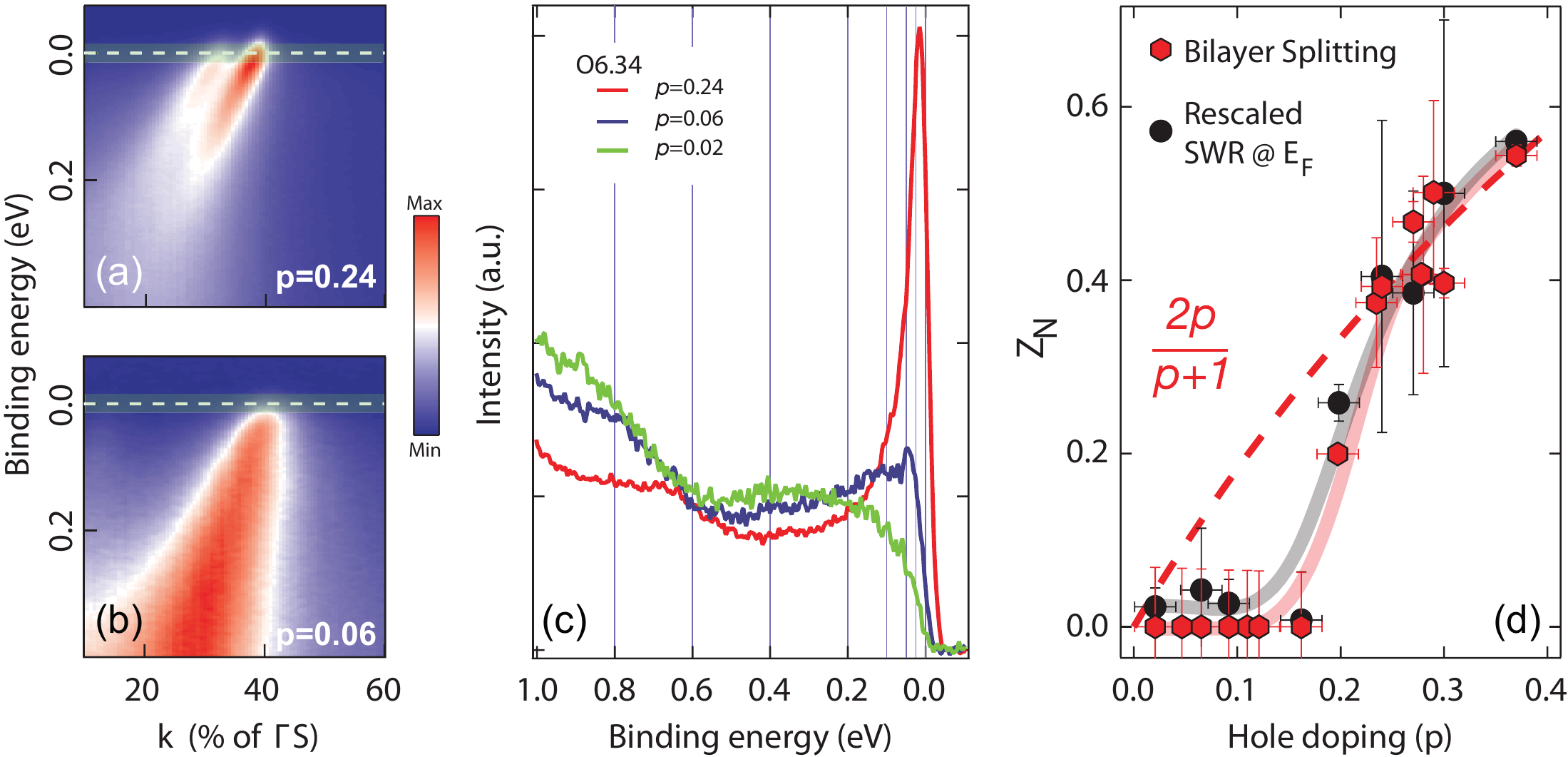,width=1\linewidth,clip=}}
\caption{(Color). (a,b): ARPES dispersion in YBCO, along the nodal cut [$ \Gamma \rightarrow (\pi,\pi) $] for $p\!=\!$ 0.24 and 0.06, respectively, showing the lack of bonding-antibonding (B-AB) bilayer splitting and the spectral function being mostly incoherent for $p\!=\!0.06$. (c) $ A (\mathbf{k}\!=\!{\mathbf{k}}_{F,N},\omega) $ as a function of doping for the bonding Cu-O band, showing the progressive suppression of the quasiparticle peak. (d) ${Z}_{N}$ as determined from the B-AB splitting and the spectral-weight ratio SWR (see text). Also shown are guides-to-the-eye and the $2p/(p+1)$ Gutzwiller projection relation (from Ref.\,\citen{Comin:David:YBCO}).}
\index{Quasiparticle weight}
\label{Comin:YBCO}
\end{figure}
\noindent
Concurrent with a modification of the Fermi surface, which evolves from large hole-like cylinders to Fermi arcs \cite{Comin:Hossain:YBCO,Comin:David:YBCO}, there is also a pronounced change in the ARPES spectral lineshape [see Fig.\,\ref{Comin:YBCO}(c), where the corresponding energy distribution curves have been extracted from the ARPES maps in panels (a,b), for $\mathrm{k}\!=\!{\mathrm{k}}_{F}$]. In particular, two major effects are observed going from the over- to the under-doped surface: (i) the progressive loss of the nodal coherent weight with no quasiparticle peak being detected at $p\!=\!0.02$ [Fig.\,\ref{Comin:YBCO}(c)], accompanied by an increase in the incoherent tail, and consistent with conservation of the total spectral weight; (ii) the suppression of the nodal bilayer splitting $\Delta {\epsilon}_{N}^{B,AB}$ shown in the ARPES intensity maps of Fig.\,\ref{Comin:YBCO}(a,b) [an even more pronounced suppression can be observed at the antinodes], which goes hand-in-hand with the redistribution of spectral weight from the coherent to the incoherent part of the spectral function. Since correlation effects suppress hopping within and between planes in a similar fashion, the renormalization of the measured bilayer splitting with respect to the prediction of density functional theory can be used as an equivalent measure of the coherent weight $Z_{\bf k}\!=\!\Delta {\epsilon}_{N}^{B,AB} / 2 {t}_{\perp}^{LDA} (N)$,  with $ {t}_{\perp}^{LDA} (N)\!\simeq\!120$\,meV. This is a more quantitative and more accurate method than estimating the spectral weight ratio between quasiparticle and many-body continuum, $ \mathrm{SWR}\!=\!{\int}_{{E}_{F}}^{- \infty} I({\mathrm{k}}_{F,N},\omega) \mathrm{d} \omega / {\int}_{0.8 eV}^{- \infty} I({\mathrm{k}}_{F,N},\omega) \mathrm{d} \omega $, since in this case the coherent and incoherent parts of $ A (\mathbf{k}, \omega) $ are not well separated. Using both methods, it is possible to observe a suppression of the coherent weight as one goes underdoped, with $Z_{\bf k}$ vanishing around $p\!=\!0.1-0.15$ [Fig.\,\ref{Comin:YBCO}(d)], consistent with the observation that the underdoped ($ p\!<\!0.1$) ARPES spectra are mostly incoherent, $ 
A (\mathbf{k}, \omega)\!\sim\!{A}_{incoh} (\mathbf{k}, \omega) $. The proximity of the Mott phase ($ p\!=\!0 $), with its strongly correlated behavior, is believed to be the reason underlying the loss of coherent behavior as hole doping is progressively reduced, and forces a departure from the Fermi liquid description much more rapid than predicted by the mean field Gutzwiller projection $Z\!=\!2p/(p+1)$.

\subsection{Temperature-controlled coherence: the manganites}
\index{Spectral weight!coherent!vs. temperature}
\index{Transition Metal Oxides!3\textit{d}-based}

%Present Mannella's work with a preliminary warning that QPs can be very tricky to find in certain materials where ${I}_{coh} \ll {I}_{incoh}$ (also here I could back-refer to Kyle's work on SCOC, and highlight similarities between the incoherent profiles). Then discuss emergence of QP/onset of coherent behavior with decreasing temperature. Maybe briefly relate this to transport properties of the material (transition between different regimes).
%
\begin{figure}[!t]
\centerline{\epsfig{figure=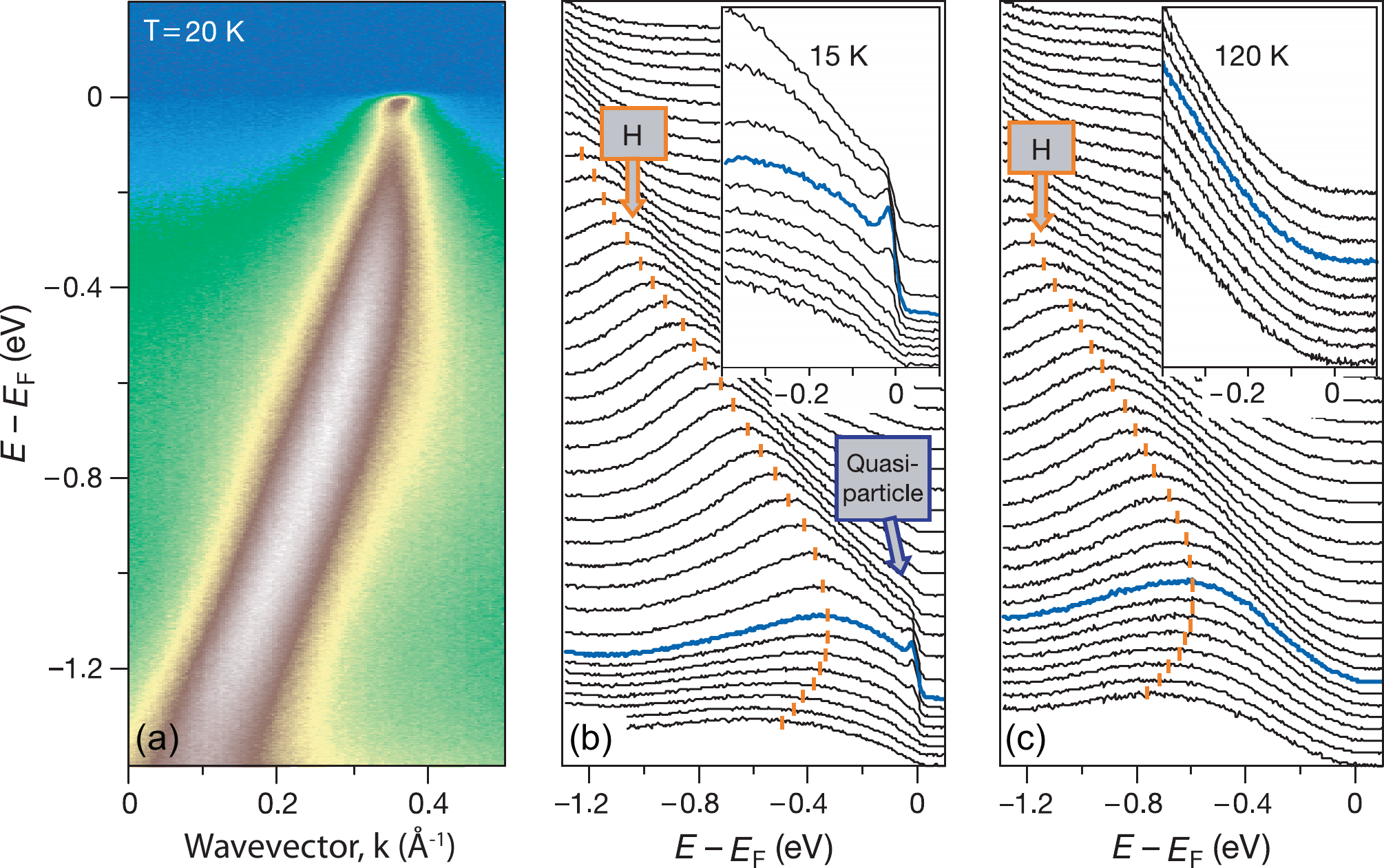,width=1\linewidth,clip=}}
\caption{(Color). (a) ARPES image plot of the Mn-${e}_{g}$ valence band along the $(0,0)$-$(\pi,\pi)$ direction in La${}_{1.2}$Sr${}_{1.8}$Mn${}_{2}$O${}_{7}$ ($T\!=\!20$\,K); note the quasiparticle band ${\varepsilon}_{k}^{q}$ branching off the bare band near ${E}_{F}$, in analogy to the case of the Holstein model [Fig.\,\ref{Comin:Polaron}(b)]. (b,c) Stack of low-energy EDCs for $T\!=\!15$ and 120\,K, respectively, emphasizing the emergence of the quasiparticle peak below ${T}_{C}$ (from Ref.\,\citen{Comin:Mannella:Manganites}).}
\label{Comin:Manganites}
\end{figure}
Another family of 3\textit{d}-based oxides characterized by a rich phase diagram is that of the manganites\index{Manganites}. These materials, which exhibit the fascinating phenomenon known as colossal magnetoresistance, have been extensively studied by ARPES \cite{Comin:Chuang:Manganites,Comin:Mannella:Manganites,Comin:Sun:PRL,Comin:Sun:NatPhys,Comin:deJong:PRB1,Comin:deJong:PRB2,Comin:Massee:NatPhys}. In one of these compounds, La${}_{1.2}$Sr${}_{1.8}$Mn${}_{2}$O${}_{7}$, the high temperature spectra ($ T\!>\!120 $\,K) do not qualitatively differ from those seen for undoped cuprates, previously presented in Fig.\,\ref{Comin:H2_model}(c). As shown in Fig.\,\ref{Comin:Manganites}(d), there is no spectral weight at the Fermi energy, and the lowest-energy excitation is a broad peak dispersing between -1 and -0.5 eV \cite{Comin:Mannella:Manganites}. This finding suggests we are again looking at a strongly correlated system, where all the spectral weight is pushed into a broad and incoherent structure away from ${E}_{F}$. Surprisingly, when the temperature is lowered through the Curie value $ {T}_{C}\!\sim\!120 $\,K, a sharp feature emerges at the chemical potential, with an intensity progressively increasing as the sample is cooled down to 15\,K [see Fig.\,\ref{Comin:Manganites}(a,b,c) and related insets]. This is a remarkable example of how temperature can lead to a transfer of spectral weight from $ {A}_{incoh} $ to $ {A}_{coh} $, in this case associated with the ferromagnetic transition occurring at $ {T}_{C} $. Note the large ratio $ {A}_{incoh} / {A}_{coh} $, i.e. a incoherent-to-coherent transition, and the subsequently small $ Z_{\bf k} $: this is an indication of the fact that we are still in a regime where electronic correlations are very strong, similar to the case of undoped and underdoped cuprates and, as we will see in the following section, also of cobaltates \cite{Comin:Wells:BiCo}. In addition, it is important to note how the coherent spectral weight does not necessarily appear throughout the entire Brillouin zone, but might instead be limited to a reduced momentum range, where electronic excitations can propagate in a coherent manner. %This might be analogous to the nodal/antinodal dichotomy in the low-energy spectral response of cuprates \cite{Comin:Damascelli:RMP}.

\subsection{Probing coherence with polarization: the cobaltates}
\index{Spectral weight!coherent!vs. photon polarization}
\index{Transition Metal Oxides!3\textit{d}-based}

%Introduce the cobaltite work by pointing out a caveat: in a multi-band system (so, unlike manganites and cuprates) the extraction of coherent and incoherent sections of SF can be awkward. A way to work this around is by means of polarization; emphasize that this approach hinges on the properties of SF (back-refer to its definition), by which its coherent and incoherent part must have the same polarization dependence. Then discuss the SF you can get via linear dichroism and compare to previous case; also discuss T-dependence (but this has less priority within this discussion).

\begin{figure}[!t]
\centerline{\epsfig{figure=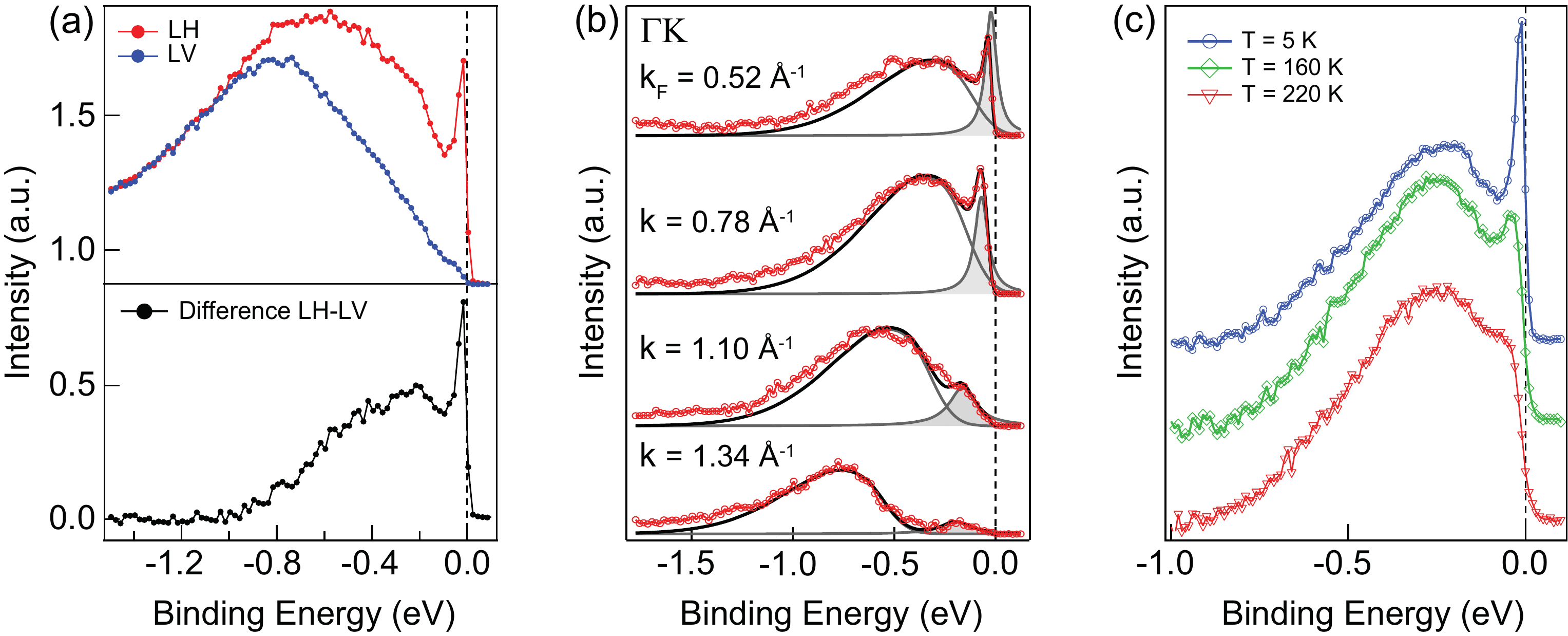,width=1\linewidth,clip=}}
\caption{(Color). (a) top panel: $ A (\mathbf{k}\!=\!{\mathbf{k}}_{F},\omega) $ from the misfit cobaltate [Bi${}_{2}$Ba${}_{2}$O${}_{4}$][CoO${}_{2}$] for two different polarizations - red curve is linear horizontal (LH), blue curve is linear vertical (LV); (a) bottom panel: linear dichroism $ {A}_{LD}\!=\!{A}_{LH} - {A}_{LV}  $. (b) momentum-dependence of $ {A}_{LD} $ near $ \mathrm{k}\!=\!{\mathrm{k}}_{F} $ (from Ref.\,\citen{Comin:Nicolaou:cobaltates}). (c) temperature-dependence of $ {A}_{LD} $ \cite{Comin:Brouet_inprogress}.}
\label{Comin:Cobaltites}
\end{figure}

As discussed in the previous sections, the distinction between coherent and incoherent parts of $ A (\mathbf{k}, \omega) $ -- and thus the determination of the quasiparticle strength $ Z_{\bf k} $ -- although conceptually well defined, is often not easy to estimate from ARPES experiments. An additional complication is encountered whenever more overlapping bands contribute to the low energy electronic structure in the same region of momentum. In such instances, there is one characteristic of the ARPES technique which can be exploited, namely the explicit dependence on light polarization of the photoemission intensity from a band of specific symmetry, as a result of matrix-element effects\index{Matrix element} (see Sec. \ref{Comin:sec:matele}). For a single band system, changing any of the experimental parameters would change the ARPES intensity as a whole, thus preserving the shape of the spectral function and in particular the ratio between $ {A}_{coh} (\mathbf{k}, \omega) $ and $ {A}_{incoh} (\mathbf{k}, \omega) $, since these terms are weighted by an identical matrix element. The situation is very different in a multiband system since, whenever the quasiparticle peaks and the many-body continua originate from different single-particle bands, they will be characterized by a different overall symmetry. In this case, one may use the polarization dependence of the single-particle matrix elements to disentangle the different spectral functions contributing to the total ARPES intensity. This approach, shown in Fig.\,\ref{Comin:Cobaltites} and discussed in more detail in Ref.\,\citen{Comin:Nicolaou:cobaltates}, has been used in the study of misfit cobaltates\index{Cobaltates}, a family of layered compounds, where the low-energy electronic states reside in the CoO${}_{2}$ planes. These compounds all have 3 bands crossing the chemical potential and, while detecting $ {A}_{coh} $ is relatively simple due to its sharpness in proximity to $ {E}_{F} $, evaluating the ratio between ${A}_{coh}$ and ${A}_{incoh}$ is a much more complicated task due the the overlap of contributions stemming from different orbitals.\\
Two close-lying bands, of respectively $ {a}_{1g} $ and $ {{e}_{g}}' $ orbital character, have different symmetries and can thus be selected using polarization, as described in Sec.\,\ref{Comin:sec:matele} [see top panel in Fig.\,\ref{Comin:Cobaltites}(a)]. Taking the difference between spectra measured with different polarization (linear dichroism), one can isolate the full spectral function for the $ {a}_{1g} $ band [Fig.\,\ref{Comin:Cobaltites}(a), bottom panel]. The momentum- and temperature-dependence are then displayed in Fig.\,\ref{Comin:Cobaltites}(b) and (c), respectively, evidencing a very similar behavior to the one found in the manganites \cite{Comin:Brouet_inprogress}. With this approach is thus possible to track the quasiparticle weight $Z_{\bf
k}$ as a function of temperature and doping, even in multiband systems and in those regions of momentum space where bands overlap.
\begin{figure}[!t]
\centerline{\epsfig{figure=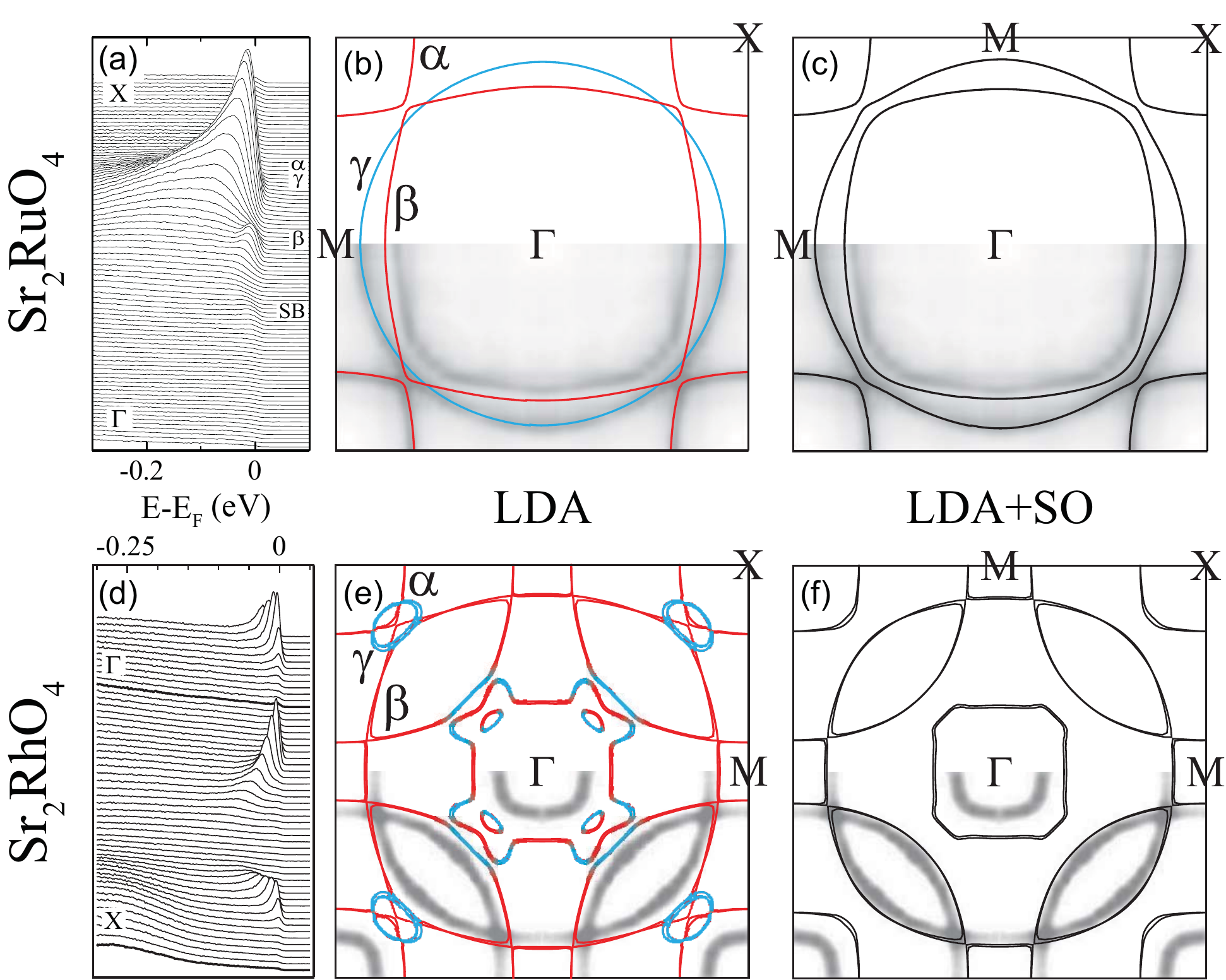,width=1\linewidth,clip=}}
\caption{(Color). (a-c): Stack of experimental ARPES EDCs in Sr${}_{2}$RuO${}_{4}$, along the high-symmetry direction $ \mathrm{\Gamma} \rightarrow \mathrm{X} $ (a), with LDA (b) and LDA+SO (c) predictions for the Fermi surface (from Ref.\,\citen{Comin:DamascelliA:Fersss,Comin:Haverkort:Sr2RuO4}). (d-f): same as (a-c), for the case of Sr${}_{2}$RhO${}_{4}$ (from Ref.\,\citen{Comin:Baumberger:Sr2RhO4,Comin:Kim:Sr2RhO4,Comin:Haverkort:Sr2RuO4}).
}
\label{Comin:4d_oxides}
\end{figure}

\subsection{Correlated relativistic metals: spin-orbit coupled 4\textit{d}-TMOs}
\index{Spectral weight!coherent!vs. band filling}
\index{Spin-orbit coupling}
\index{Transition Metal Oxides!4\textit{d}-based}

%Now turn page, and justify the scientific case of looking into heavier d-based oxides, to play with the underlying energy scales (maybe worth back-referring to some previously introduced hierarchy of scales in 3d materials, or to the Mott criterion). Discuss two novel aspects: (i) correlations are weaker, metallic states emerge even in stochiometric compounds; suddenly U is not so dominant, this changes the landscape dramatically (refer to sharp QP pole in EDCs in Fig.\,\ref{Comin:4d_oxides}) (ii) simple-LDA methods fail to account for the topology of the Fermi surface (refer again to figure). Use the latter point to introduce role of SOC (point out magnitude for the cases of Rh and Ru)
Stepping down one row in the periodic table we find the 4\textit{d} transition metal oxides. Based on simple arguments, one would expect correlations to play a less important role in these materials. This is due to the larger spatial extent of the 4\textit{d} orbitals, as compared to the 3\textit{d} case, which at the same time favors delocalization (larger \textit{W}) and reduces on-site electron-electron interactions (smaller \textit{U}), thus positioning these systems away from the \textit{Mott criterion}. Following such intuitive expectations, one indeed finds an evident suppression of correlation effects, which is accompanied by the emergence of coherent charge dynamics even in undoped (i.e., stoichiometric) compounds. However, marking the difference from 3\textit{d} oxides, a new important term has to be considered for 4\textit{d} materials: the spin-orbit (SO) interaction. The associated energy scale $ {\zeta}_{SO} $ becomes increasingly important for heavier elements (with an approximate $ {\zeta}_{SO}\!\propto\!{Z}^{4} $ dependence on the atomic number \textit{Z}), which then have to be treated within a relativistic framework. Whereas these effects are largely neglected in cuprates, where ${\zeta}_{SO} ({\mathrm{Cu}}^{2+})\!\sim\!20-30$\,meV, they are important in ruthenates and rhodates (and even more in 5\textit{d} materials, as we will see later), where ${\zeta}_{SO} ({\mathrm{Ru}}^{4+})\!=\!161$\,meV and ${\zeta}_{SO} ({\mathrm{Rh}}^{4+})\!=\!191$\,meV \cite{Comin:Earnshaw:4d}. Furthermore, in 4\textit{d} systems correlation effects continue to play a role, hence these systems are commonly classified as \textit{correlated relativistic metals}.

ARPES results on two of the most studied 4\textit{d}-based oxides, namely Sr${}_{2}$RuO${}_{4}$ and Sr${}_{2}$RhO${}_{4}$ \cite{Comin:DamascelliA:Fersss,Comin:Baumberger:Sr2RhO4,Comin:Kim:Sr2RhO4,Comin:Haverkort:Sr2RuO4} are shown in Fig.\,\ref{Comin:4d_oxides}, together with predictions for the Fermi surface from density functional theory in the local density approximation (LDA)\index{Ruthenates!Sr${}_{2}$RuO${}_{4}$}\index{Rhodates!Sr${}_{2}$RhO${}_{4}$}\index{Quasiparticle}. While one indeed finds intense and sharp quasiparticle peaks in the energy distribution curves (EDCs) -- and consequently large values of $Z_{\bf
k}$ supporting the strongly reduced relevance of many-body correlations -- the matching between experimental and predicted Fermi surfaces is not perfect for Sr${}_{2}$RuO${}_{4}$ and is actually poor for the even more covalent Sr${}_{2}$RhO${}_{4}$. Experiments and theory are almost fully reconciled when SO coupling is included in the single-particle methods used to describe the low-energy electronic structure of 4\textit{d}-oxides; on the other hand, the experimental bands still appear renormalized with respect to the calculations, by approximately a factor of 2 similar to overdoped cuprates \cite{Comin:David:YBCO,Comin:Plate:Tl2201}, which indicates that electronic correlations cannot be completely neglected. This ultimately qualifies Sr${}_{2}$RuO${}_{4}$ and Sr${}_{2}$RhO${}_{4}$ as \textit{correlated relativistic metals}.

\subsection{Mott criterion and spin-orbit coupling: 5\textit{d} TMOs}
\index{Spin-orbit coupling}
\index{Spectral weight!coherent!vs. spin-orbit interaction}
\index{Transition Metal Oxides!5\textit{d}-based}

\begin{figure}[!t]
\centerline{\epsfig{figure=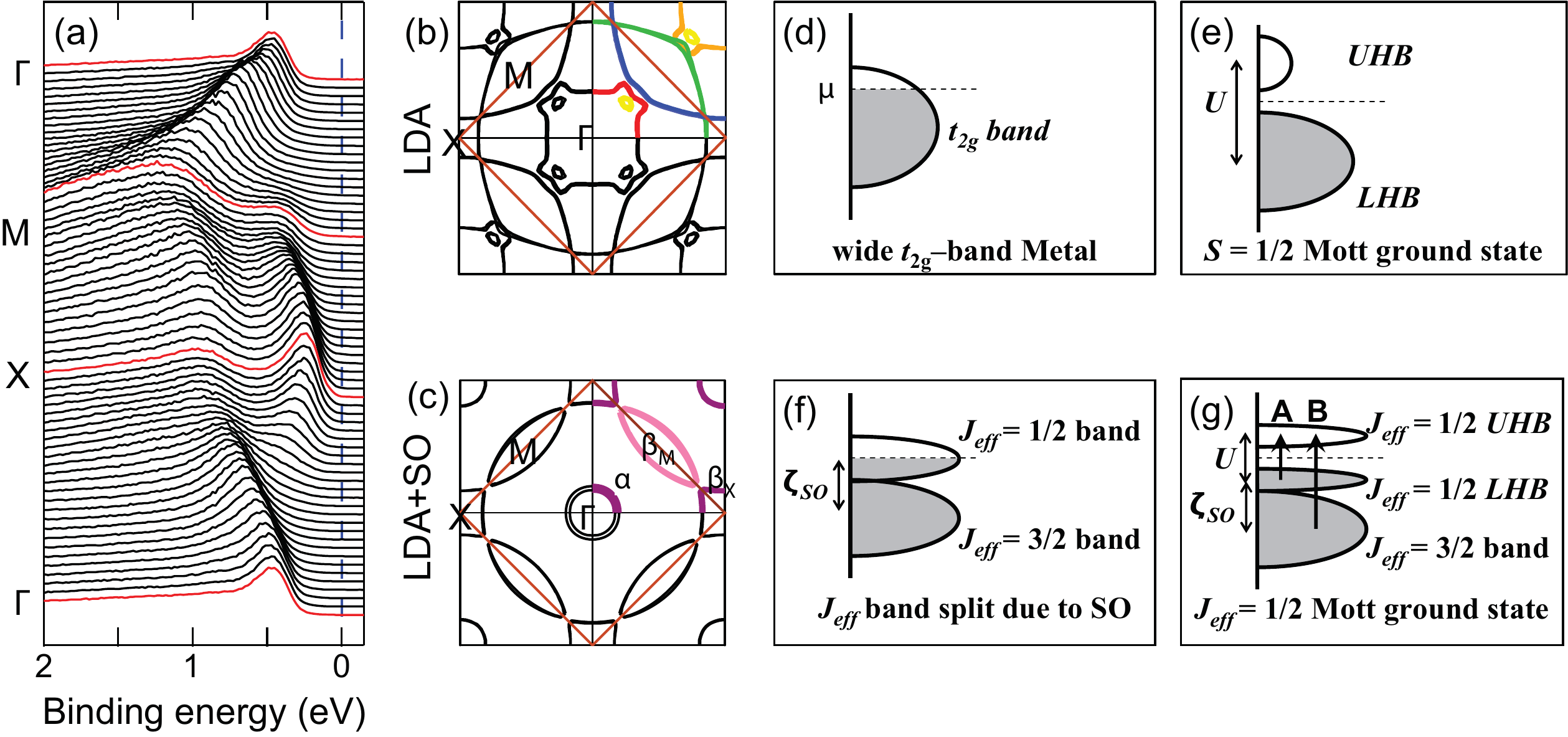,width=1\linewidth,clip=}}
\caption{(Color). (a) ARPES data along high-symmetry directions within the first Brillouin zone (EDCs at high-symmetry points are marked in red), showing no spectral weight at $ {E}_{F} $ (blue dashed line). (b,c): density-functional calculations within the LDA and LDA+SO approximations, respectively. (d,e) Possible low-energy scenarios in a $ 5{d}^{5} $ system: (d) $U\!=\!0$, ${\zeta}_{SO}\!=\!0$, yielding an uncorrelated metallic ground state; (e) $U\!>\!W$, ${\zeta}_{SO}\!=\!0$, yielding a $ S\!=\!1/2 $ Mott-insulating state; (f) $U\!=\!0$, ${\zeta}_{SO}\!\sim\!W$, giving a spin-orbit coupled metal; (g) ${\zeta}_{SO}\!\sim\!W$, $U\!\sim\!{W}_{{J}_{eff}=1/2}$, producing a $ {J}_{eff}\!=\!1/2 $ Mott-insulating ground state (from Ref.\,\citen{Comin:Kim:Sr2IrO4}).}
\label{Comin:5d_oxides}
\end{figure}

%Introduce the 5d-oxides by arguing about expectations of even lower correlated behaviour than 4d-materials, and also highlight their novelty (1 decade history for these compounds). Maybe quote a few conflicting works, or give a hang of the variety of phenomena predicted, to justify the idea that these materials are inherently complex. Then show experimental results on Sr${}_{2}$IrO${}_{4}$ (Fig.\,\ref{Comin:5d_oxides}) and the findings of insulating behavior in the first place. Discuss Kim's work and the original proposal of Mott-insulating state in the strong-SOC limit for (but still leave it sound like an open problem).
Based on the reduced correlation effects observed in 4\textit{d}-oxides, a progressive evolution into an even less correlated physics in 5\textit{d} materials would be expected. For this reason, the discovery of an insulating state in Sr${}_{2}$IrO${}_{4}$, a compound isostructural and chemically similar to cuprates and ruthenates, came as a big surprise. The first resistivity profiles to be measured in this iridate \cite{Comin:Cao:Sr2IrO4}\index{Iridates!Sr${}_{2}$IrO${}_{4}$} showed an insulating behavior, as also later confirmed by optical spectroscopy \cite{Comin:Kim:Sr2IrO4}. ARPES data on this material, showing the low-energy dispersions of the Ir 5\textit{d}-${t}_{2g}$ states, consistently found no spectral weight at ${E}_{F}$ [see Fig.\,\ref{Comin:5d_oxides}(a)]. 
\noindent
Furthermore, and most importantly, there is a significant disagreement between experimental data and LDA(+SO) calculations that, as displayed in Fig.\,\ref{Comin:5d_oxides}(b,c), would predict the system to be metallic, with a Fermi surface corresponding to a large Luttinger counting. This is a situation reminiscent of the 3\textit{d}-oxides, where fulfilment of the \textit{Mott criterion}\index{Mott!criterion} would yield a correlated $ S\!=\!1/2 $ insulating state at variance with band theory. This novel underlying physics emerges because of the prominent role of the SO interaction, whose strength is $ {\zeta}_{SO}\!\sim\!500$\,meV for Ir${}^{4+}$, and which now acts in concert with the other relevant energy scales (\textit{W} and \textit{U}). In the atomic limit, the action of SO would splits the otherwise degenerate ${t}_{2g}$ orbitals into two submanifolds with the total angular momentum $ {J}_{eff}^{2}\!=\!{L}_{eff}^{2}+{S}^{2} $, and its projection $ {J}_{eff}^{z} $, as new quantum numbers. Within such a framework, local correlations would then split the $ {J}_{eff}\!=\!1/2 $ manifold into lower and upper Hubbard bands thus opening a Mott gap, provided $ U\!>\!{W}_{{J}_{eff}=1/2} $. This mechanism, sketched in Fig.\,\ref{Comin:5d_oxides}(f,g), yields a novel type of correlated ground state, the so-called \textit{relativistic Mott-insulator}\index{Insulator!Mott relativistic}.
One should note that the validity of such \textit{pseudospin-1/2} approximation is still debated, and alternative mechanisms are being discussed. Also, recent works have questioned the Mott-like nature of the electronic ground state in Sr${}_{2}$IrO${}_{4}$ and rather suggested that this system could be closer to a \textit{Slater-type} (thus, non-correlated) insulator \cite{Comin:Arita:Sr2IrO4,Comin:Hsieh:Sr2IrO4}. In this latter case, the insulating gap would result from the onset of long-range magnetic ordering and not strong electron correlations of the Mott type, i.e. the metal-insulator transition would coincide with the magnetic ordering transition. In the next and last section, we will present an unambiguous experimental realization of relativistic Mott physics in iridates, and will highlight similarities and differences with respect to Sr${}_{2}$IrO${}_{4}$.

\subsection{Relativistic Mott insulating behavior: Na${}_{2}$IrO${}_{3}$}
\index{Transition Metal Oxides!5\textit{d}-based}

%Discuss the importance of evaluating the relevant energy scales in iridates, in order to establish a valid effective in these complex systems, which are at the boundary of strongly (3d) and weakly (4d) correlated behavior (so-that no strong- or weak-coupling limit can be used to simplify treatment). Present Na${}_{2}$IrO${}_{3}$ results as first work to reveal that in these materials all energy scales have to be treated on equal footing. Emphasize the importance of combining multiple spectros=copies to properly characterize these materials. Outline possible perspectives for future investigations on these materials.
%
\begin{figure}[!t]
\centerline{\epsfig{figure=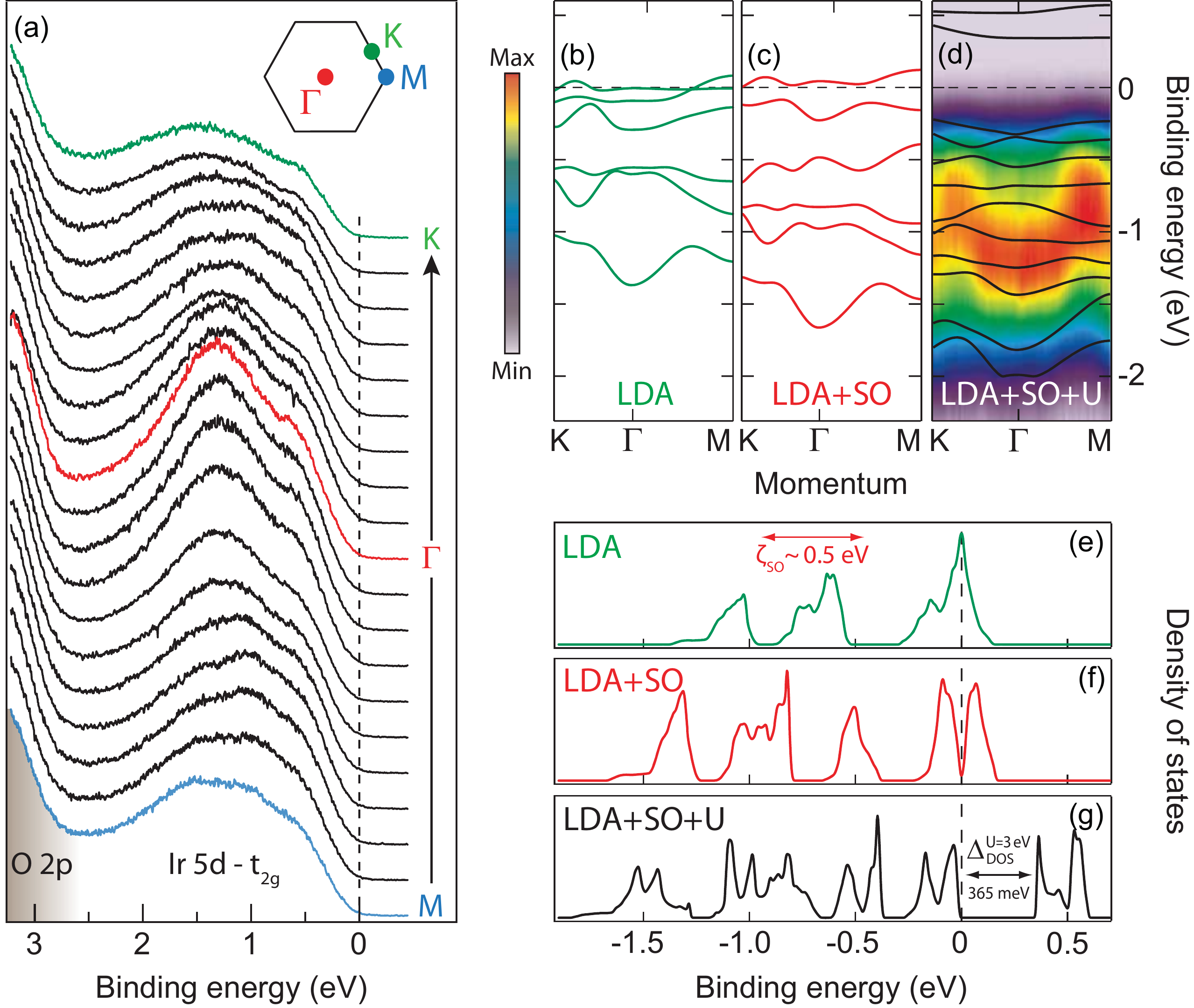,width=1\linewidth,clip=}}
\caption{(Color). (a) Experimental EDCs from ARPES, along the high-symmetry directions $ \mathrm{M} \rightarrow \mathrm{\Gamma} \rightarrow \mathrm{K} $ (as defined in the hexagonal surface-Brillouin zone, top-right corner). Ir 5\textit{d} states populate the range 0-2\,eV, whereas the O 2\textit{p} states are found at ${E}_{B}\!>\!2$\,eV. (b,c,d) DFT band structure from LDA, LDA+SO, and LDA+SO+U [with ARPES intensity map superimposed in (d)]. (e,f,g) Corresponding theoretical densities of states (DOS) for the various LDA methods (from Ref.\,\citen{Comin:Comin:Na2IrO3}).}
\label{Comin:Na2IrO3}
\end{figure}
After the original discovery and proposal of a Mott-insulating state in Sr${}_{2}$IrO${}_{4}$, new systems were predicted, both on theoretical and experimental grounds, to exhibit similar physics. Na${}_{2}$IrO${}_{3}$\index{Iridates!Na${}_{2}$IrO${}_{3}$} is one of these compounds, which has been the subject of early theoretical speculations \cite{Comin:Shitade:Na2IrO3,Comin:Jin:Na2IrO3,Comin:Chaloupka:Na2IrO3}, and was later synthesized and found to behave in a correlated manner. Early transport and magnetization measurements \cite{Comin:Singh:Na2IrO3} provided evidence for an insulating behavior characterized by local spin moments, therefore pointing to a Mott scenario (charge localization). Further spectroscopic evidence for such a scenario has been provided by a combination of ARPES, optics and LDA calculations \cite{Comin:Comin:Na2IrO3}. ARPES data for the Ir 5\textit{d}-${t}_{2g}$ bands from a pristine surface are shown in Fig.\,\ref{Comin:Na2IrO3}(a), and highlight a few primary aspects: no spectral weight is found at $ {E}_{F} $ and the band dispersions are surprisingly narrow ($W\!\sim\!0.15$\,eV). Moreover, the energy distribution curves are very broad, with no evidence of sharp quasiparticles, perhaps as a result of a vanishing ${Z}_{\mathbf{k}}$, unlike the case of Sr${}_{2}$IrO${}_{4}$. While this is unexpected for a system possessing the more extended 5\textit{d} orbitals, it also brings us closer to fulfilment of the Mott criterion $ U\!>\!W $ and therefore to a correlated, Mott-Hubbard-like physics. LDA predicts this system to be metallic with a density of states peaking at ${E}_{F}$ [see Fig.\,\ref{Comin:Na2IrO3}(b,e)]. The disagreement with the ARPES data implies that the charge dynamics cannot be explained within a simple band-model.  When also accounting for the SO interaction, it is found that this is sufficient to turn the system insulating within the LDA calculations [Fig.\,\ref{Comin:Na2IrO3}(c)], although with a 0-gap at ${E}_{F}$ [Fig.\,\ref{Comin:Na2IrO3}(f)]. It is only with the further inclusion of the Coulomb term \textit{U} that the correct gap size ${\Delta}_{gap}\!\sim\!340$\,meV, as found from optics and ARPES with potassium evaporation \cite{Comin:Comin:Na2IrO3}, can finally be reproduced [see Fig.\,\ref{Comin:Na2IrO3}(d,g), where values of $U\!=\!3$\,eV and ${J}_{H}\!=\!0.6$\,eV have been used]. The presence of a sizeable on-site electron-electron interaction also explains the presence of local moments well above the long-range antiferromagnetic ordering temperature ${T}_{N}\!\sim\!13$ K (i.e., in the paramagnetic phase). These findings reveal that the expectation of correlation-free physics in selected 5\textit{d} oxides is in general unrealistic, and that spin-orbit coupling has a primary role in making these systems unstable against possibly small correlation effects (in this case, the on-site Coulomb interaction). This indicates that many-body and spin-orbit interactions cannot be fully disentangled, thus conclusively establishing Na${}_{2}$IrO${}_{3}$ -- and possibly other members of the iridates family -- as \textit{relativistic Mott insulators}: i.e., a novel type of correlated insulator in which many-body (Coulomb) and relativistic (spin-orbit) effects have to be treated on an equal footing.

\vfill

%\bibliographystyle{spphys}
%\bibliography{Springer_ARPES}

\providecommand{\noopsort}[1]{}\providecommand{\singleletter}[1]{#1}%

% BibTeX users please use
% \bibliographystyle{}
% \bibliography{}
%
% Non-BibTeX users please follow the syntax
% the syntax of "referenc.tex" for your own citations
%\input{referenc}
%%%%%%%%%%%%%%%%%%%%%%%%%%%%%%%%%%%%%%%%%%%%%%%%%%%%%%%%%%%%%%%%%%%%%%  }

%%%%%%%%%%%%%%%%%%%%%%%%%%%%%%%%%%%%%%%%%%%%%%%%%%%%%%%%%%%%%%%%%%%%%%

\printindex
\end{document}